\documentclass[fleqn,usenatbib]{mnras}

\usepackage{newtxtext,newtxmath}
\usepackage[T1]{fontenc}
\usepackage{ae,aecompl}

\usepackage{graphicx}	
\usepackage{amsmath}	
\usepackage{amssymb}	

\usepackage{natbib}
\usepackage{amsmath}
\usepackage{graphicx}
\usepackage[colorinlistoftodos]{todonotes}
\usepackage{xcolor}

\newcommand{\name}{DeepSource~}

\newcommand{\namediff}{DS$_D$}
\newcommand{\namesame}{DS$_S$}

\newcommand{\snr}{signal-to-noise ratio}
\usepackage{float}

\usepackage{todonotes}

\title[DeepSource9]{DeepSource: Point Source Detection using Deep Learning}

\author[A. Vafaei Sadr et al.]{A. Vafaei Sadr, $^{1,2,3}$\thanks{E-mail: vafaei.sadr@gmail.com}
Etienne. E. Vos, $^{2,4,5}$\thanks{E-mail: etienne.vos@ibm.com}
Bruce A. Bassett, $^{2,3,5,6}$\thanks{E-mail: bbassett@ska.ac.za}
\newauthor
Zafiirah Hosenie,$^{2,5,7}$
N. Oozeer,$^{2,3}$ and 
Michelle Lochner$^{2,3}$
\\
$^{1}$  Department of Physics, Shahid Beheshti University, Velenjak, Tehran 19839, Iran\\
$^2$ African Institute for Mathematical Sciences, 6 Melrose Road, Muizenberg, 7945, South Africa\\
$^3$ South Africa Radio Astronomical Observatory, The Park, Park Road, Pinelands, Cape Town 7405, South Africa\\
$^4$ IBM Research Africa, 45 Juta Street, Braamfontein, Johannesburg, 2001, South Africa \\
$^5$ South African Astronomical Observatory, Observatory, Cape Town, 7925, South Africa\\ 
$^6$ Department of Maths and Applied Maths, University of Cape Town, Cape Town, South Africa\\
$^7$ Jodrell Bank Centre for Astrophysics, School of Physics and Astronomy, The University of Manchester, Manchester M13 9PL, UK\\
}

\date{Accepted XXX. Received YYY; in original form ZZZ}

\pubyear{2018}

\begin{document}
\label{firstpage}
\pagerange{\pageref{firstpage}--\pageref{lastpage}}
\maketitle
\begin{abstract}
Point source detection at low signal-to-noise is challenging for astronomical surveys, particularly in radio interferometry images where  the noise is correlated. Machine learning is a promising solution, allowing the development of algorithms tailored to specific telescope arrays and science cases. We present \name - a deep learning solution - that uses convolutional neural networks to achieve these goals. \name enhances the Signal-to-Noise Ratio (SNR) of the original map and then uses dynamic blob detection to detect sources. Trained and tested on two sets of $500$ simulated $1^\circ \times 1^\circ$ MeerKAT images with a total of $300,000$ sources, \name is essentially perfect in  both purity and completeness down to SNR = $4$ and outperforms PyBDSF in all metrics. For uniformly-weighted images it achieves a {\em Purity} $\times$ {\em Completeness} (PC) score at SNR = $3$  of $0.73$, compared to $0.31$ for the best PyBDSF model. For natural-weighting we find a smaller improvement of $\sim 40\%$ in the PC score at SNR = $3$. If instead we ask where either of the purity or completeness first drop to $90\%$, we find that \name reaches this value at SNR = $3.6$ compared to the $4.3$ of PyBDSF (natural-weighting). A key advantage of \name is that it can learn to optimally trade off purity and completeness for any science case under consideration. Our results show that deep learning is a promising approach to point source detection in astronomical images. 
\end{abstract}



\section{Introduction}~\label{sec:Introduction}

Creating a catalogue of point sources is an important step in the science exploitation of most astronomical surveys, yet none of the existing algorithms is satisfactory at low Signal-to-Noise Ratio (SNR) for radio interferometers \citep{Hopkins}. Traditionally, in radio astronomy, one extracts the noise (rms; $\sigma$) from a ``source free'' region \citep{2000MNRAS.315..808S, 2013MNRAS.435..650M} and then identifies sources as those peaks above a threshold \citep{2018A&C....23...92C}, often taken to be 3$\sigma$, depending on the image quality. Next a Gaussian would be fit to the source and the respective parameters extracted. This was the strategy of the original Search And Destroy (SAD) algorithm, an AIPS~\footnote{Astronomical Image Processing Software (AIPS) - http://www.aips.nrao.edu/index.shtml}. Although source finding has evolved and several variants developed, most of these methods suffer from the problem that the choice of the location and size of the region used for estimating the noise is susceptible to contamination from faint point sources which can lead to biases in source counts due to false positives and false negatives, especially at SNR $< 5$, but sometimes even at a SNR $< 10$ (see \cite{Hopkins}).  

Moreover, the big data challenges posed by upcoming massive astronomical facilities such as the Square Kilometre Array (SKA) will require robust and fast source detection algorithms to leverage the full potential of the large images that will be produced. An  example of the challenges that will arise is that the point spread function (PSF) of the image will now vary across the wide fields and broad bandwidths of the observations. Such complications can be tackled by BIRO, a sophisticated Bayesian forward model for the entire system, including nuisance parameters that encodes potential systematic errors (such as pointing offsets) and fits all parameters directly in the visibilities 
~\citep{2015MNRAS.450.1308L}. This has the advantage of being statistically rigorous and principled. However it is computationally expensive and not yet developed for practical radio interferometry purposes. 

An alternative is to look to machine learning to improve source detection, inspired by the human-level performance of deep learning in a wide variety of applications \citep{LeCun_2015, He_2015} and is quickly becoming prevalent in astronomy; a few examples include  the simulation of  galaxy images \citep{Ravanbakhsh}, estimation of photometric redshifts \citep{Hoyle}, performing star-galaxy classification \citep{Kim}, transient classification \citep{Trans1,Trans2} and unsupervised feature-learning for galaxy Spectral Energy Distributions (SEDs)~\citep{Frontera-Pons}. In the context of point source detection there are several conceptual advantages of machine learning over traditional approaches, including the possibility of optimising the point source detection for a given science case, by trading off completeness for purity.  Deep learning is a particularly promising avenue for source detection since, once trained, the algorithms are very fast and hence well-adapted to handle the petabytes of data that will be flowing from new radio sky surveys.

In this work, we use deep learning in the form of Convolutional Neural Networks (CNNs)~\citep{LeCun1995, Krizhevsky_2, schmidhuber2015deep}. The main advantage of a CNN is that it automatically learns representative features from the data directly, for example in images it recovers high level features from lower lever features. On the contrary, traditional machine learning or pattern recognition algorithms require manual feature extraction to perform prediction. CNNs do not make any prior assumption on specific features but rather automatically learn the relevant descriptors directly from the pixels of the images during training.

We present a deep Convolutional Neural Network for point source detection called \name, that is able to learn the structure of the correlated noise in the image and thus is able to successfully detect fainter sources. We compare \name to the Python Blob Detection Source Finder (PyBDSF, formerly known as PyBDSM~\cite{Mohan}), one of the most robust point source detection algorithms. Our primary aim is to benchmark the robustness and performance of \name compared to PyBDSF by assessing the relative completeness and purity of the two algorithms. We have limited this work to only point sources but this approach can further be explored for more complex source structures.

The layout of this paper is as follows. In Section \ref{Model Simulation}-\ref{Image Generation}, we present the model simulation and the data we use for this work. 
A detailed overview of \name is presented in Section \ref{sec:DeepSource}. The results for the different cases are illustrated and discussed in Section \ref{Results}.

\section{Model Simulation}\label{Model Simulation}

To begin with, we simulated input catalogues of point sources and images for the MeerKAT (64 $\times$ 13.5 m dish radio interferometer) telescope, which is one of the SKA precursors~\citep{2007mru..confE...7J}. 

\begin{figure}
\centering
\includegraphics[width=0.49\textwidth]{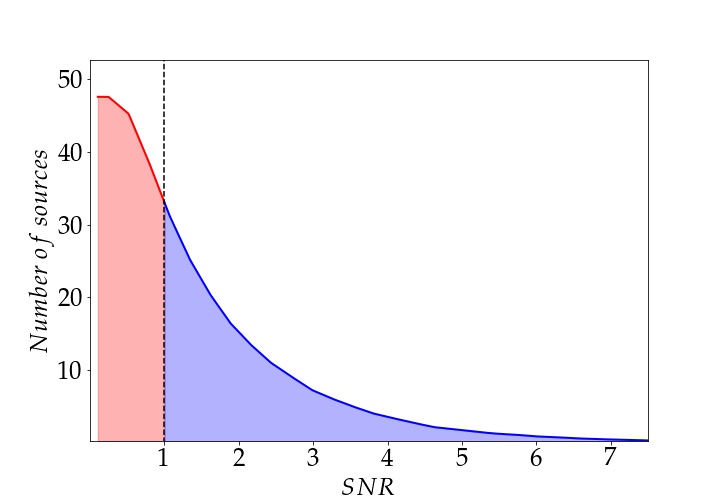}
\caption{\label{fig:exponential_distribution} The average distribution of the number of sources in bins of width 0.033, with respect to Signal to Noise Ratio (SNR) over 500 sets of simulated images. We chose to sample from an exponential distribution and since we are mostly interested in faint source detection we selected parameters to ensure that $\approx$ 45\% of the sources to have SNR $<$ 1.0 with a cut-off at SNR $=$ 0.1.}
\end{figure}

We sampled the SNR from an exponential distribution with a cut-off at SNR $= 0.1$, which approximates the expected luminosity function, Fig.~\ref{fig:exponential_distribution}.  We chose our parameters such that $\approx$ 45\% of the sources have SNR $<$ 1.0. SNR was computed by taking the ratio of the pixel-flux of a particular source/detection divided by the noise of the image.  The pixel-flux refers to the flux value of the pixel located at a given set of coordinates for a ground-truth source or a detection coming from \name or PyBDSF.  The noise of the image, which is defined as the standard deviation of the background flux, is calculated from a ``source-free'' region of the image border where by construction there are no sources (see the border of Fig. \ref{fig:images}). To compute the flux, a normalization constant of $\mathrm{10 \times 10^{-8}}$ Jy was used. 

We created two sets of catalogues: which we call {\em Same-Field} and {\em Different-Field}, each with 500 images of size $1^\circ \times 1^\circ$. Each image in the catalogues has 300 randomly distributed sources per square degree. The {\em Different-Field} images each have different field centres, drawn from $  10.0^{\circ} < RA < 110.0^{\circ} $ and $ -70.0^{\circ} < DEC < 30.0^{\circ}$. The {\em Same-Field} simulations on the other hand all have the same field centres (RA = $50.0^{\circ}$ and DEC = $-20.0^{\circ}$). The simulated position of the sources were randomly assigned in RA and DEC ensuring a separation of at least $(\approx 40.0'')$ which is $4$ times the synthesized beam of the MeerKAT telescope in the high-resolution maps. Though this means that there are no overlapping sources in the high-resolution case, this was not the case in the low-resolution maps. Finally, the point sources were assigned intrinsic major axis and minor axis length of $3''$, significantly smaller than the beam in all cases and hence ensuring that all sources are effectively point-like.

\subsection{Image Generation}\label{Image Generation}
\begin{table}
\label{Parameters for MeerKAT images}
\caption{Parameters used to simulate each MeerKAT sky image. In addition we simulated two groups of images: {\em Different-Field}, defined by images whose centres were drawn randomly on the sky between $10^{\circ} < {\rm DEC} < 110^{\circ}$ and $-70^{\circ} < {\rm RA} < 30^{\circ}$ and {\em Same-Field} images, all of which were centered at ${\rm RA} = 50.0^{\circ}$ and ${\rm DEC} = -20.0^{\circ}$. }
\begin{center}
	\begin{tabular}{cc}
    	\hline\hline
        {Parameters} & Value\\
        \hline\hline
        Synthesis time& $1$ h\\
        Exposure time& $5$ min\\
        Frequency of Observation& $1400$ MHz\\
        Channel Width& $2.0898$ MHz\\
        Number of Channel& $10$\\
        System Equivalent Flux Density (SEFD) & $831$ Jy\\
        Aperture Efficiency& $0.7$\\
        System Temperature& $30$ K\\
        Noise in visibilities& $1.0\times 10^{-4}$ Jy\\
        Number of Sources simulated & $300$\\
        Pixel scale& $1.5''$\\
        Image size  & $4096\times4096$ pixels\\
        Image size  & $1.25^{\circ} \times 1.25^{\circ}$\\
        Robustness Parameter, R& $(-1.5, 1.5)$\\
        Clean iterations& $10 000$\\
        Auto-threshold& $3.0$\\

        \hline
	\end{tabular}   
\end{center}

\end{table}

We simulated MeerKAT sky images using Stimela\footnote{ \href{https://github.com/SpheMakh/Stimela/wiki}{https://github.com/SpheMakh/Stimela/wiki}, a pipeline that combines most radio interferometry tools into a Docker container: https://www.docker.com/.} which at its core uses the Meqtrees package based on the measurement equations \citep{meqtrees}. The simulated images have a pixel size of 1.5 arcsecond. Using the MeerKAT configuration as given in Table \ref{Parameters for MeerKAT images} as our array, we generated  Measurement Sets (MS) consistent with the generated catalogues described previously.  

\begin{figure*}
\centering
\includegraphics[width=\textwidth]{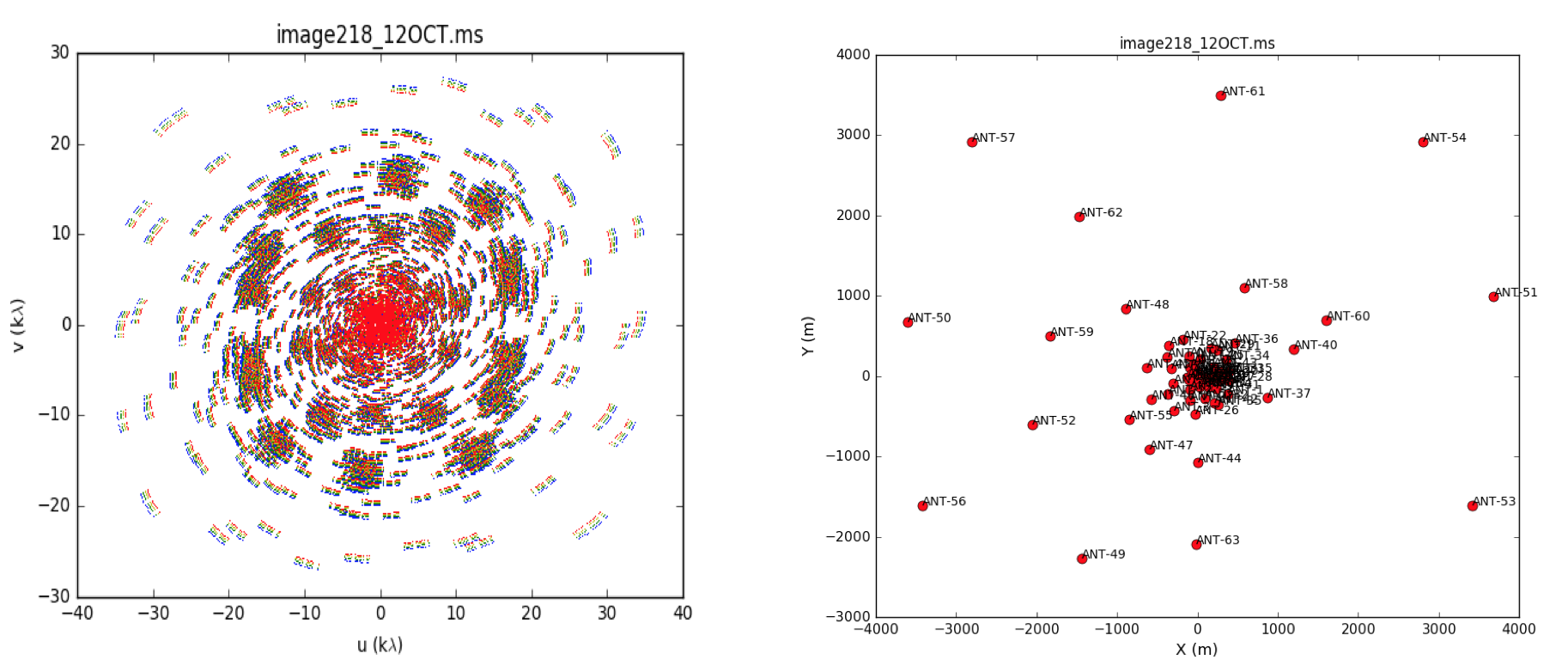}
\caption{\label{fig:u-v} A sample uv-coverage plot (left) for a simulated observation of 1 hour synthesis time with 5 minutes exposure time. The spatial layout of the 64 MeerKAT  antennae is also shown (right), illustrating the dense core.}
\end{figure*}

The MeerKAT channel width from the continuum correlator is around 208.984 kHz. In our simulation, we chose to average 10 channels (making a channel width of 2.08984 MHz). To more realistically simulate MeerKAT observations, the $u-v$ coverage was sampled from a noise-free sky for a simulated 1 hour synthesis time and 5 min exposure times with the full 64 antenna configuration.  A noise of $1.0 \times 10^{-4} Jy$ was then added to the visibilities in the $u-v$ plane in the XX and YY polarizations. 
The choice of these parameters were based on the expected rms we wanted to achieve. A sample $u-v$ of a one hour observation is shown in Fig.~\ref{fig:u-v} for one of our simulations using the antenna positions from MeerKAT.

Once the MS of the visibilities has been created, it is then passed to the WSClean algorithm \citep{Offringa} for deconvolution. WSClean is an astronomical widefield imaging algorithm, based on the original CLEAN algorithm \citep{Hogbom}, for radio interferometric data. WSClean is powerful, fast and well adapted for our analysis. 

\subsection{Weightings}\label{sec:weightings}

One of the main problems of imaging is the incomplete $u-v$ coverage that needs to be corrected for in any interferometer. This is achieved by deconvolving the Point Spread Function (PSF) of the instrument in the images. Poor imaging comes from the fact that we have limited a priori knowledge of the sky brightness distribution as well as the  large gaps in the $u-v$ coverage. This challenge can be addressed in various ways. One is during the imaging step, where each visibility is given a weight. The weighting helps to account for the noise variances in different samples and to enhance sensitivity of extended sources. In our simulation, we have implemented a flexible weighting scheme developed by \cite{Briggs} that smoothly varies a function of a single real parameter, the Robustness, $R$ \citep{Briggs}. 

The two weighting schemes we use are: 

\textbf{Natural Weighting}: This weights all visibilities equally and results in maximum point-source sensitivity in images. However, this weighting provides a poor synthesized beam-shape and side-lobe levels. It corresponds to $R = 1.5$.

\textbf{Uniform Weighting}: This gives the visibilities a weight that is inversely proportional to the sampling density function. This causes a reduction in the side-lobes of the PSF. The uniform weighting scheme provides better resolution but lowers the sensitivity of the image. It corresponds to $R = -1.5$.

\begin{figure*}
\begin{center}
\includegraphics[width=0.47\textwidth]{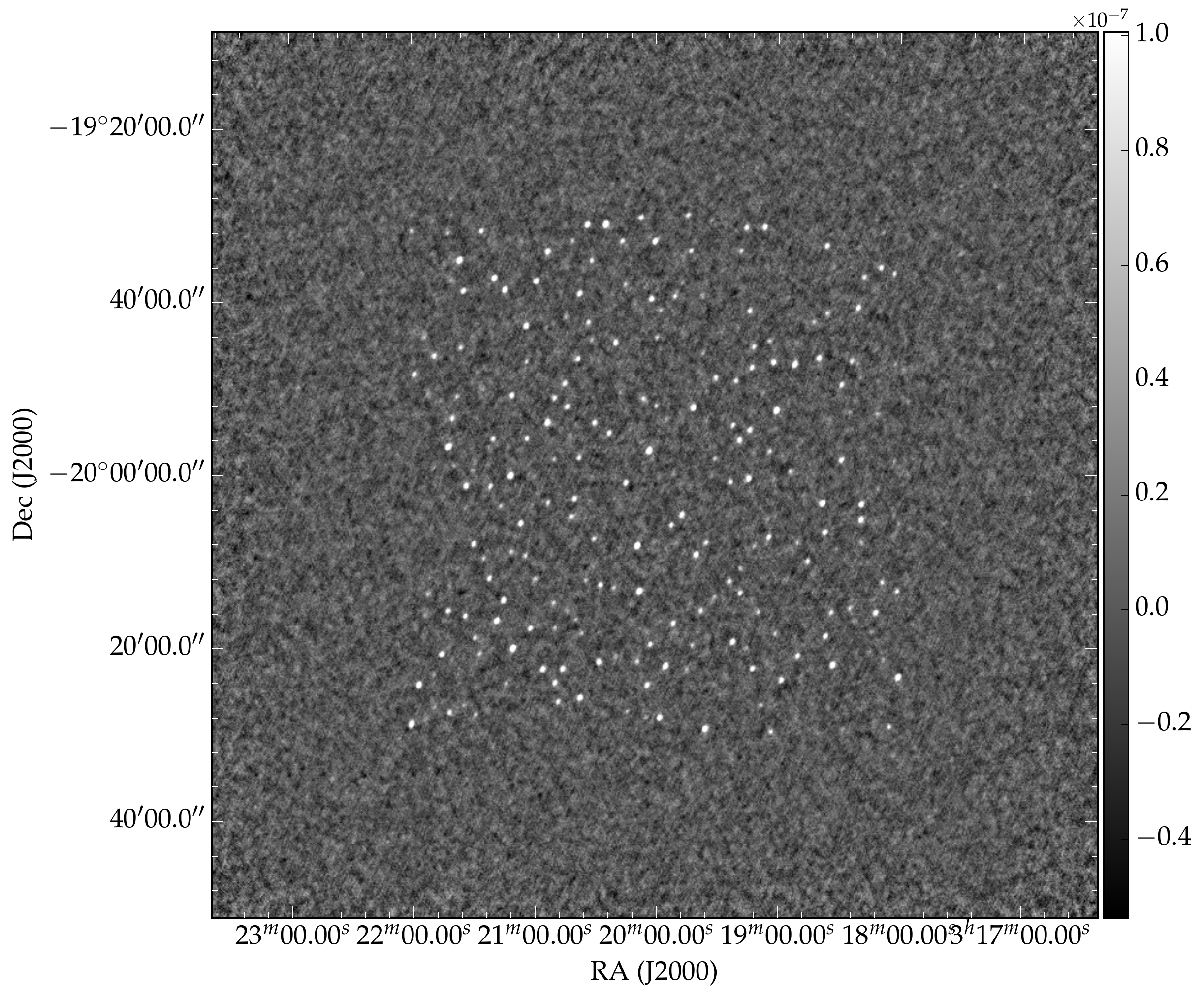}
\includegraphics[width=0.47\textwidth]{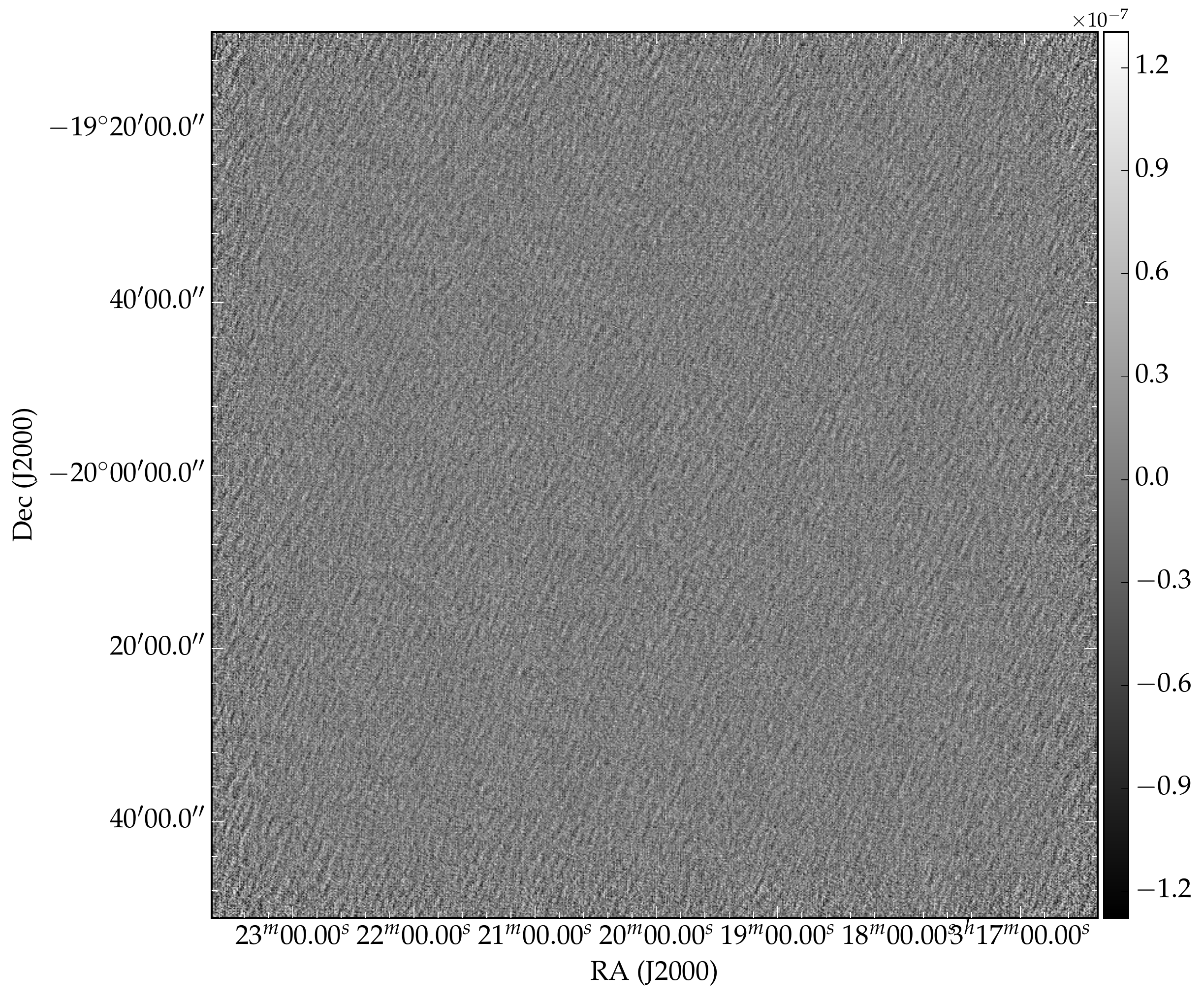}
\caption{ \label{Images_of_different_weightings} Two simulated images created with natural {\em (Left)} and uniform weightings {\em (Right)} from the same input source model. The images include sources distributed randomly having at least 4 $\times$ the synthesized beam separation between each other with flux density distribution sampled from an exponential distribution. 
}
\label{fig:images}
\end{center}
\end{figure*}

For every simulated observation, we created two images with $R = 1.5$ (natural weighting) and $R = -1.5$ (uniform weighting). Fig.~\ref{Images_of_different_weightings} shows two simulated images for the same input model. The noise in natural-weighted images is $\approx 15$ nJy while for uniform-weighted images it is  $\approx 40$ nJy. 

The imaging of the data was carried out on a 40-core server with 4 GPUs. A fixed number of iterations (10 000) was applied for the deconvolution step for each images. As a consequence, the noise in the image simulations are  higher than the theoretical noise. 

The synthesized beam is around $10''$ and we used a  pixel size of $1.5''$ with $4096 \ \times \ 4096$ pixels. To avoid edge effects in the images, and to allow accurate estimation of the noise, we included border regions without any point sources around the central core of each image, see e.g. Fig.~\ref{fig:images}. 

As mentioned in Section \ref{Model Simulation}, we have ensured that each source is separated by at least $40''$ from any other source. After simulating the images, we calculated the beam size of the different robustnesses for a single image as follows:

\begin{equation} \label{Beam_size}
\textrm{Beam size} = \sqrt{B_{\rm maj} \times B_{\rm min}}
\end{equation}

where $B_{\rm maj}$ and $B_{\rm min}$ are the synthesized beam major and minor axis in arcseconds. The natural-weighted image has a beam size of $\approx 27.0''$ while the uniform-weighted image has a beam size of $\approx 6.0''$. The final image size was $~1.25 \times 1.25\;\textrm{deg}^{2}$ including the source-free border. 

\section{PyBDSF}
\label{PyBDSF} 

As discussed in Section \ref{sec:Introduction}, many source detection packages are available (see \cite{Hopkins}). We compared our analysis with PyBDSF, one of the state-of-the-art algorithms. PyBDSF finds islands of neighbouring emission in the images by locating all pixels that are higher than a user defined pixel threshold. Each of these pixels is then added to an island of contiguous pixels exceeding the island threshold.  Prior to source detection, PyBDSF computes the background rms noise and the mean of the image which later creates islands of a given user-defined sigma level. PyBDSF then fits one or multiple Gaussians to each detected source. 

A $2\sigma$ island threshold was used to determine the location in which source fitting is done. We also used a $2\sigma$ peak threshold such that only islands with peaks above this threshold are added to the catalogue. These thresholds were chosen because they gave the best results on a validation set, ensuring the fairest possible comparison with PyBDSF. 

The resulting Gaussians are finally grouped to produce a list of  candidate sources from the image. The Gaussian fitting produced various fitted parameters including the position of the maximum, the peak value of the maximum, the width of the Gaussian and the position angle etc. Detailed information about PyBDSF and its various algorithm available can be found in the documentation\footnote{\href{http://www.astron.nl/citt/pybdsm/index.html}{http://www.astron.nl/citt/pybdsm/index.html}.}.

\section{DeepSource}\label{sec:DeepSource}

Our approach is based on Convolutional Neural Networks (CNN), driven by the expectation that CNN should be able to extract information buried in the correlated noise in order to better separate sources from the background, thus effectively increasing the sensitivity of the telescope. 

Deep learning, i.e. neural networks with many layers, has lead to great improvements in performance over the past decade. CNNs are typically applied to classification problems where the image is mapped to a single class label. In addition, CNNs are also the method of choice for various other computer vision related tasks such as image segmentation, object recognition and localization \citep{richardson2013learning,shen2015deepcontour}.

In the following sections we introduce \name - a CNN-based method that learns to separate point sources from correlated background noise.  We also discuss the algorithm and key concepts that form the core of \name.

\subsection{The \name Algorithm}

A CNN is a set of filters that are arranged in consecutive layers in a neural network.  These filters represent the learnable weights of the network and are usually initialized randomly at the beginning of training.  The network changes the weights in order to minimize a predefined loss, which is a function that measures the difference between the network output and some known target output (referred to as the ground truth).  A CNN is essentially a highly optimized, nonlinear transformation between a given input and an expected output.

In this study we use simulated noisy images of the radio sky containing point sources at different positions for the input to the CNN, and corresponding ground-truth maps that contain only the point sources as the target output.  In the ground truth maps, the locations of point sources are given by single-pixel delta functions, so that all pixels are set to zero except for those that correspond to the true point source locations.  This image is referred to as the $\delta$-image.  An example of input image and $\delta$-image for a single point source is given in the left and right panels of Figure~\ref{point_source}, respectively.

The CNN can essentially be regarded as a transformation from the input image $\mathcal{I}$ to a demand image $\mathcal{D}$, where each point source in the demand image is represented by a single-pixel delta function (as given by the $\delta$-image) at the true locations of the point sources which are available from the simulations of the noisy images.  With this approach, the goal is that the CNN learns to amplify regions that contain ``real'' point sources and to suppress correlated noise elsewhere.

Ideally one would want to transform $\mathcal{I}$ (which may include multiple channels) to the delta-function image. However there are several limitations which drive $\mathcal{D}$ away from the $\delta$-image. To see these limitations let us choose an appropriate loss function.

A natural choice of loss function for our problem would be the sum of the squares of the difference between the CNN output $\mathcal{T}$ and the desired demand image $\mathcal{D}$:
\begin{figure}
\begin{center}
\includegraphics[width=0.22\textwidth]{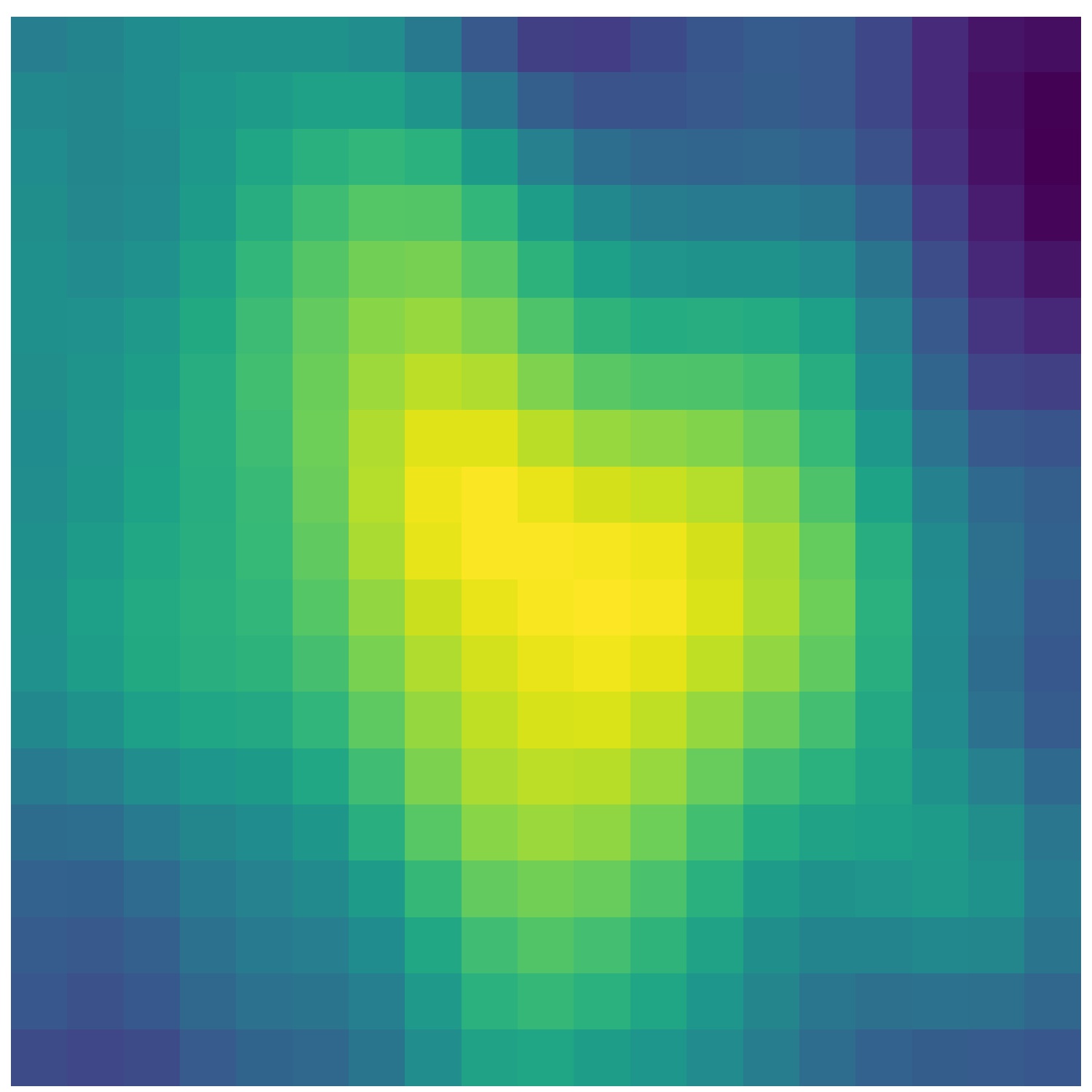}
\includegraphics[width=0.22\textwidth]{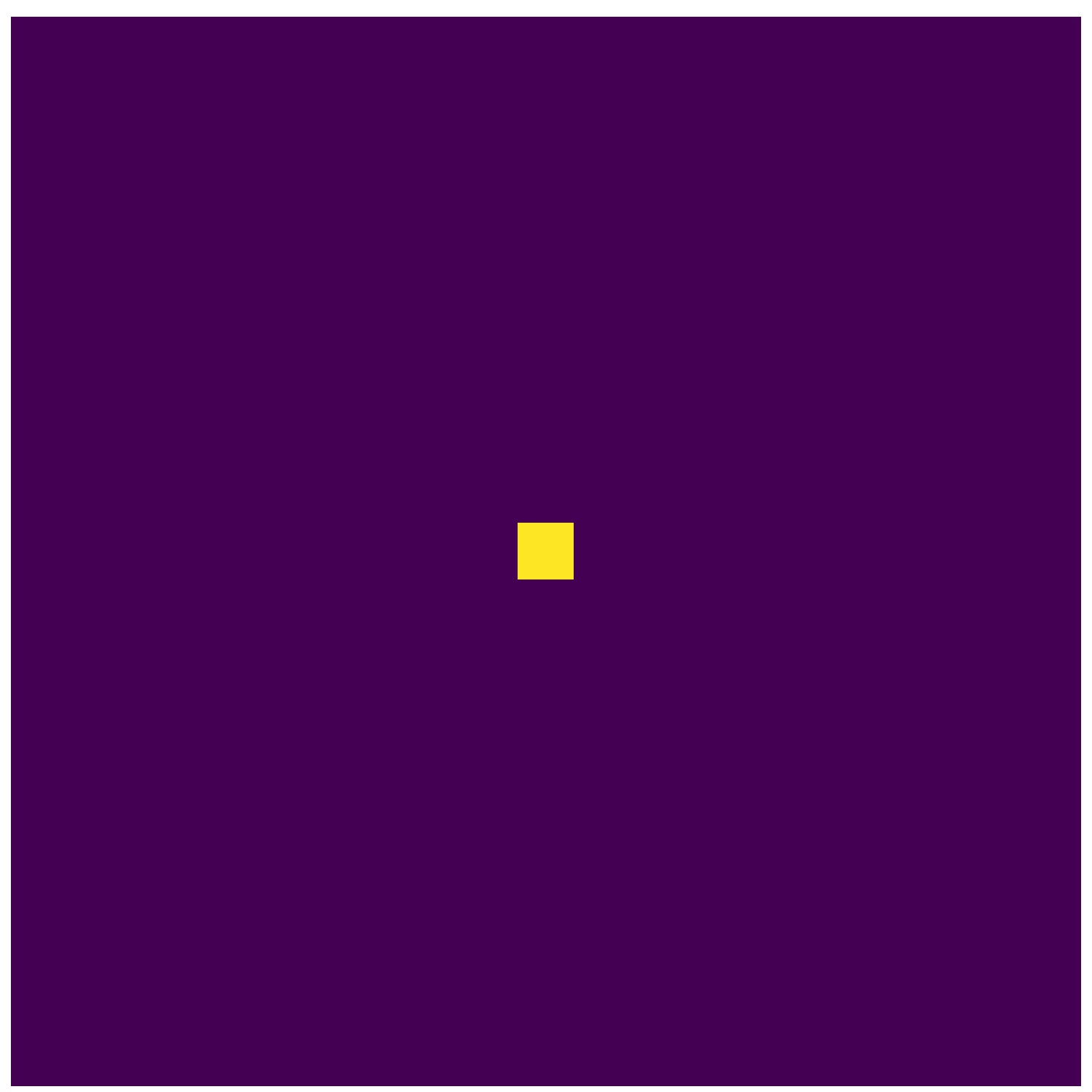}
\end{center}
\caption{Left: an actual image of a point source with undesired effects (e.g. beam and noise). Right: what is expected to be seen in an ideal observation. An ideal transformation would suppress all of the undesired effects and transform the left image to the right one. }
\label{point_source}
\end{figure}

\begin{equation}
\label{loss_function}
\mathcal{L}=\Sigma_i \left(\mathcal{T}_i-\mathcal{D}_i\right)^2.
\end{equation}

If the $\delta$-image is used to represent point sources in $\mathcal{D}$, the zero-valued pixels greatly outnumber the non-zero pixels by a factor of $10^5:1$.  This leads to difficulty in training the CNN to minimize the loss $\mathcal{L}$, and in practice result in the CNN producing zero-valued pixels everywhere as the optimal solution.  A possible interpretation for this could be that, when using only single-pixel delta functions to represent point sources, the semantic information provided by pixels surrounding a point source is effectively ignored.  Representing point-sources as single-pixel delta functions in $\mathcal{D}$ is therefore sub-optimal.

A workaround to this problem is to increase the importance of the regions surrounding a point source by including more information from these regions in the output map $\mathcal{D}$.  Practically, this is achieved by smoothly varying the pixel values of $\mathcal{D}$ from zero (for locations far away from a source) to one (at the location of a source).  Smoothing the $\delta$-image in such a way helps the CNN to better learn about the vicinities in which point sources are located.

In this work we considered two sequential 2D smoothing functions:  a {\em horn-shaped} smooth triangle function and a Gaussian kernel.  The effective sizes of the smoothing functions are hyper-parameters so that we have to find the best values for them through trial and error. These play an important role in both the training process and the accuracy of the results.  A very small length makes $\mathcal{L}$ inefficient for training, while a very large length makes finding the point source location inaccurate. Our choices of the smoothing scale for the horn-shaped and Gaussian windows are $15$ and $10$ pixels respectively.

Another subtlety that we had to account for is the fact that point sources have different intensities.  The CNN can easily learn to preferentially detect bright sources compared to faint sources which are buried in the noise, even though we are interested in detecting both bright and faint sources. On the other hand, equal contributions to the loss function from faint and bright sources would confuse the CNN and raise the false positive rate. Therefore we have to decrease the faint point source contribution in comparison to bright sources. This way, the CNN learns about bright sources first and learns about fainter ones in subsequent training steps. This means we have another hyper-parameter to control: the contribution of each point source according to its brightness. A relatively large contribution of faint sources makes CNN keen to dig deeper for their detection. However it raises the noise contribution as well, therefore increasing the false positives.  

Considering all of the above, we use the following pixelwise equation to define the values of any pixel $\mathcal{D}_{ij}$:
\begin{equation}
\mathcal{D}_{ij} =(\mathcal{C}_{ij}+b_n)\mathcal{I}^{\alpha}_{ij}
\end{equation}
where the coefficient image, $\mathcal{C}_{ij}$, is the $\delta$-image smoothed by the kernels described above. The hyper-parameter $\alpha$ in the exponent of $\mathcal{I}$ controls the contribution to the loss function from point sources with different \snr. Intuitively, choosing $\alpha=0$ would cause all point sources to appear equally bright in $\mathcal{D}$. This forces the CNN to search for all point sources, including those that are buried deep inside the noise and difficult to retrieve. On the other hand, setting $\alpha$ too large will cause the CNN to focus only on the bright sources. The parameter $\alpha$ therefore controls the intensity contribution of the different sources and regulates the similarity between point sources in the demand image and those in the real image. 
A principled approach would be to choose $\alpha$ such that it follows the underlying luminosity function of the sources. However, for generality we simply optimize $\alpha$ as a hyper-parameter.

We also introduced the parameter $b_n$, which is multiplied by the  pixels in the input image $\mathcal{I}$.  This parameter preserves a small portion of $\mathcal{I}$ in $\mathcal{D}$ and forces the CNN to learn about the background. Since $b_n$ causes the background $\mathcal{D}$ to be non-zero, it helps the CNN to perform well even with small smoothing scales (which is preferable since it leads to more accurate localisation of point sources).  The values of $\alpha$ and $b_n$ are obtained with hyper-parameter estimation (see Section \ref{optim}).

This CNN-based approach benefits from several advantages. Firstly the CNN can in principle learn to ignore specific properties such as beam effects, correlated noise or other contamination.  Further, since it includes only convolutional layers without shrinking the image size, the CNN can operate on images or windows of any size. In other words, one can train it on $200 \times 200$ pixel images and then use it on $4000 \times 4000$ pixel images. Finally, it is computationally efficient and can run in parallel or on GPUs using appropriate neural network packages.
 
\subsection{Thresholded Blob Detection}\label{object_detection}

\begin{figure*}
\begin{center}
\includegraphics[width=\textwidth]{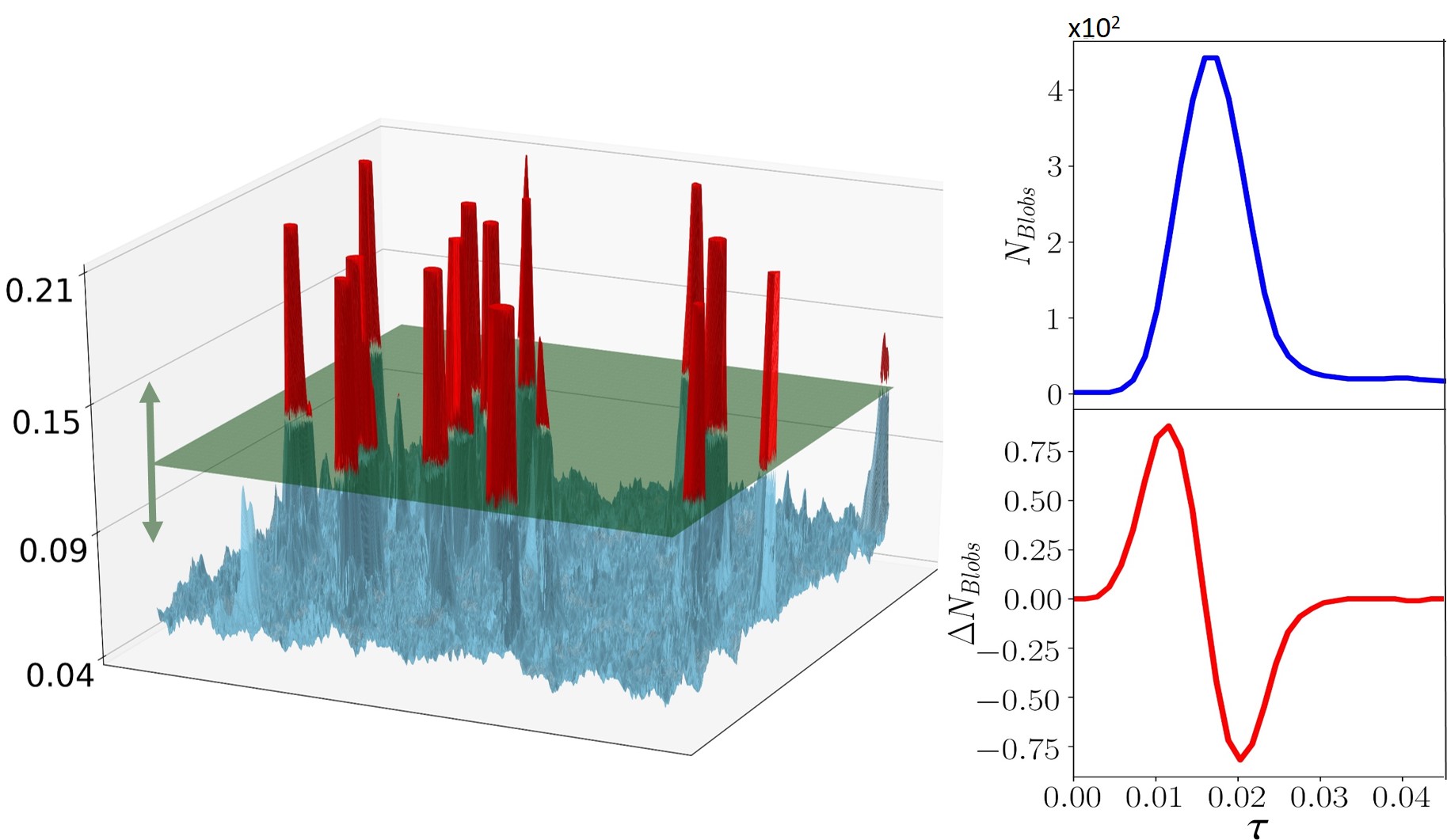}
\end{center}
\caption{The Thresholded Blob Detection (TBD) procedure. CNN output is passed to a dynamic threshold finder to find the best threshold and identify blobs in the image. The left image is a 3D view of the CNN output separated by a threshold (the plane) into light blue and red regions. The regions below the threshold  (light blue) are ignored and the regions above the threshold (red) lead to blobs that are potential point source regions. The right upper figure shows how the number of blobs, $N_{\rm blobs}(\tau)$, varies against different thresholds. The right bottom panel shows $\Delta N_{\rm blobs}=N_{\rm blobs}(\tau_{i+1})-N_{\rm blobs}(\tau_i)$ at each step of the threshold variation.}
\label{tbd}
\end{figure*}

The CNN described above effectively increases the signal-to-noise-ratio of the original image, and can therefore be considered as forming part of the pre-processing stage.  In order to obtain a catalog of predicted point source locations we introduce a technique called Thresholded Blob Detection (TBD).  The left panel of Figure \ref{tbd} shows how thresholding can produce several separate islands or blobs.  TBD uses dynamic thresholding to find the optimal threshold $\tau_{*}$.

Let us assume that the number of blobs at a given threshold $\tau$ is $N_{\rm blobs}(\tau)$. As shown in the upper-right panel of Figure \ref{tbd}, decreasing $\tau$ from the maximum to minimum values of $\mathcal{T}$($\mathcal{I}$) smoothly increases the number of blobs that appear above the threshold, up until a point where the blobs begin to merge, which leads to a smooth decrease in $N_{\rm blobs}$.
We use $\Delta N_{\rm blobs}(\tau)=N_{\rm blobs}(\tau_{i+1})-N_{\rm blobs}(\tau_i)$ (where $i$ denotes the $i^{th}$ step) to obtain the optimal value for $\tau$. For this purpose we define the maximum allowed $\Delta N_{\rm blobs}(\tau)$ as $\Delta_{max}$. The algorithm selects the $\tau_*$ by decreasing $\tau$ until $\Delta N_{\rm blobs}(\tau)$ is as close as possible to $\Delta_{max}$

At this point each blob is a potential point source. Since we are not often interested in large $\tau$ values and we need more accuracy for smaller $\tau$ values (lower SNR) we used logarithmic steps for computational efficiency. $\Delta_{max}$ is determined through hyper parameter optimization (see Section \ref{optim}). Blobs that are much smaller than the beam of the telescope are likely to be noise fluctuations. We therefore introduce $A_{min}$, which is a hyper-parameter that refers to the minimum allowed area for a blob to be considered as a source. 

Once we have identified a putative point source using TBD, the source center needs to be determined. We consider three different definitions for this:

\begin{equation}
\label{mean}
\textbf{r}_{\rm mean}=\int_{\rm blob} \textbf{r}^{\prime} d\textbf{r}^{\prime}
\end{equation}

\begin{equation}
\textbf{r}_{\rm centroid}=\int_{\rm blob} I(\textbf{r}^{\prime}) \textbf{r}^{\prime} d\textbf{r}^{\prime}
\end{equation}

\begin{equation}
\textbf{r}_{\rm peak}=r(I_{\rm max})
\end{equation}
where $\textbf{r}$ denotes pixel location in the image and $I(\textbf{r})$ is the intensity of the CNN output at pixel $\textbf{r}$.  
The optimal definition is also considered to be a hyper-parameter. We discuss  hyper-parameter optimization next in Section~\ref{optim}.

\begin{figure*}
\begin{center}
\includegraphics[width=\textwidth]{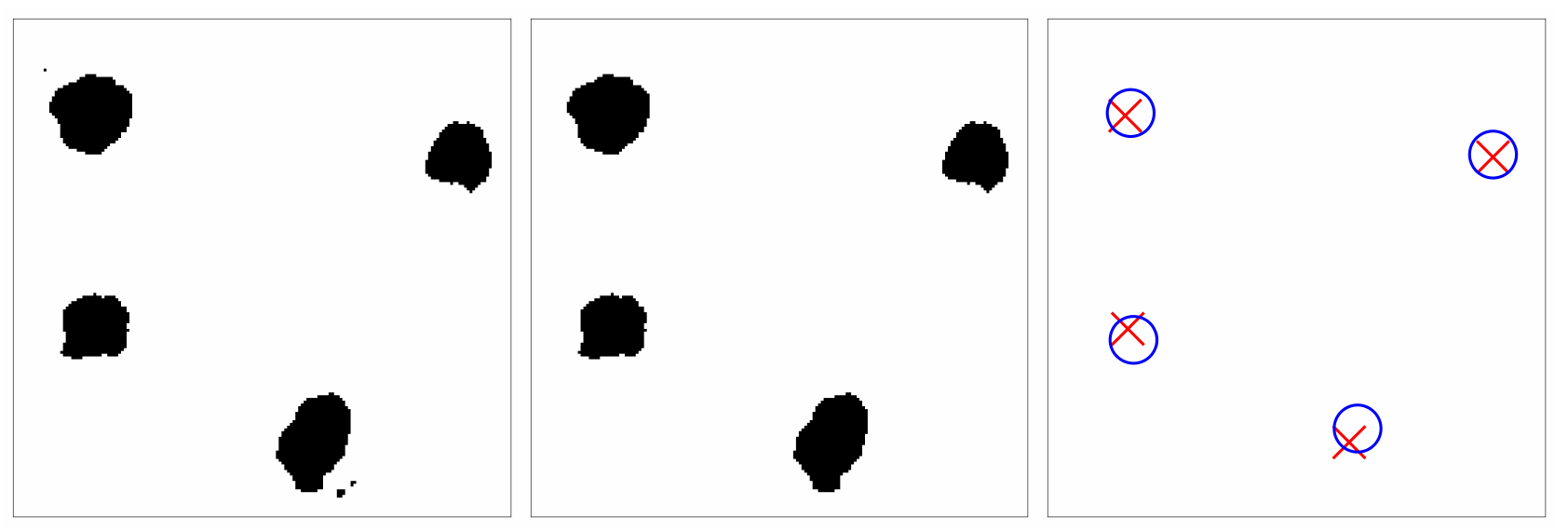}
\end{center}
\caption{A schematic view of the role of the $A_{\rm min}$ hyper-parameter that removes small blobs whose area is less $A_{\rm min}$. Left: the seven original blobs with amplitudes above the threshold $\tau$. Note that three of the blobs are very small and unlikely to correspond to real sources. Middle: small outlier blobs removed, reducing the number of blobs decreased to $4$. Right: the extracted point sources, found using equation \ref{mean}, shown by red crosses with ground truth locations show by blue circles.}
\label{detection}
\end{figure*}

\begin{figure*}
\begin{center}
\includegraphics[width=\textwidth]{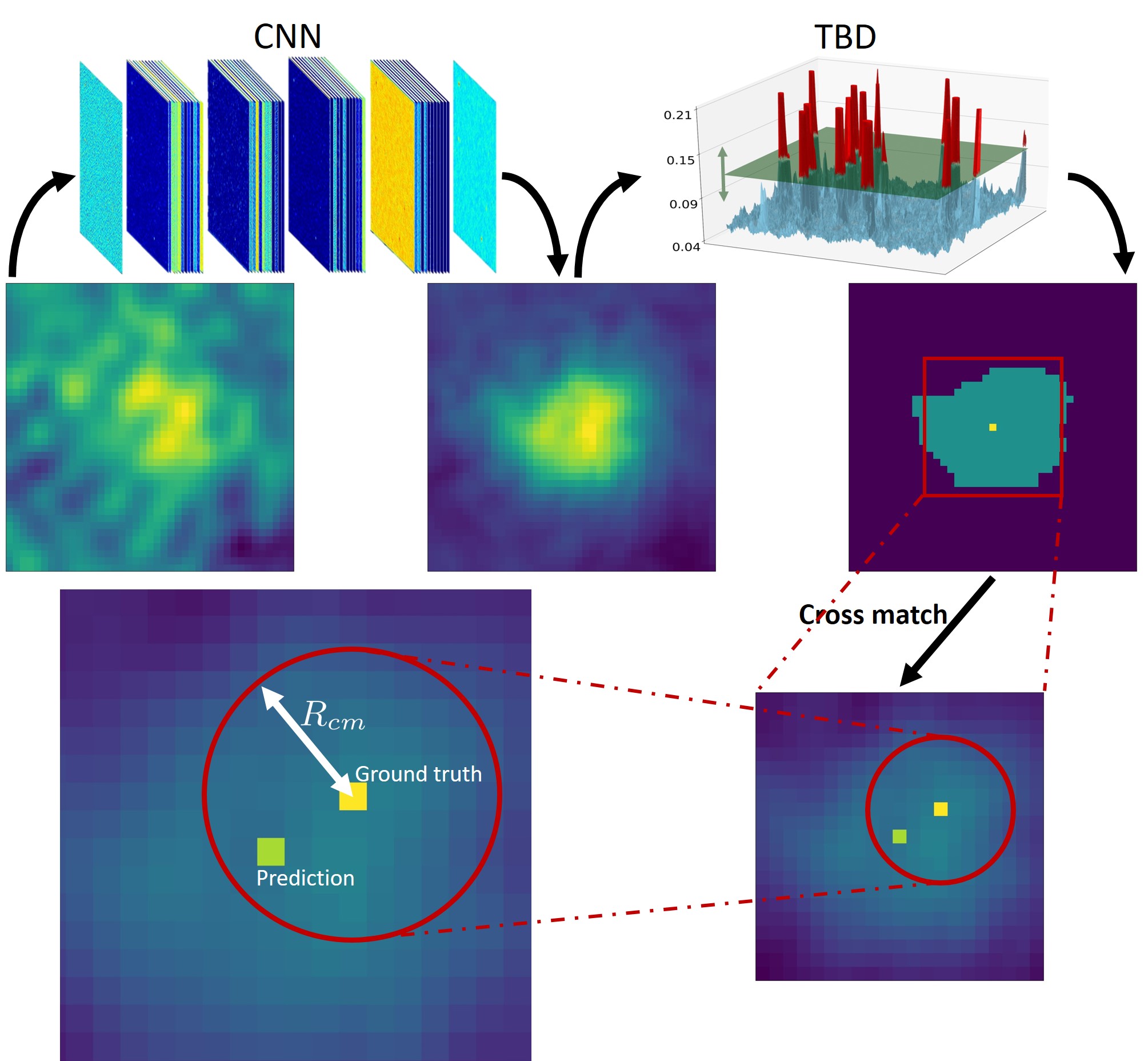}
\end{center}
\caption{Flowchart of the detection strategy of \name. On the upper left patch in the flowchart, image $\mathcal{I}$ (here one point source) is passed to the CNN and the result is an increased signal-to-noise-ratio image, $\mathcal{T}(\mathcal{I})$, shown in the middle. Then Thresholded Blob Detection (TBD) finds the best threshold, detects blobs and assigns one point source to each blob. An example of the blob and detected point source is shown in the upper right patch. Finally the detected point source is cross matched with ground truth, as shown in lower part.}
\label{flow}
\end{figure*}

\subsection{Hyper-parameter optimization}\label{optim}

\begin{table*}
\begin{center}
	\begin{tabular}{c|c}
    	\hline\hline
        Hyper-parameter & Description \\
        \hline
        smoothing scale & controls contribution of point sources vicinities' information\\
        $\alpha$ & controls contribution of different SNR sources\\
        $b_n$ & controls contribution of background information\\
        $\Delta_{\rm max}$ & stops threshold finder\\
        $A_{\rm min}$ & removes outlier blobs \\
        center definition & extracts blob centers\\
        \hline\hline
	\end{tabular}   
    \caption{A description of the various hyper-parameters used in \name.}
\end{center}
\end{table*}

In \name  we need to find the hyper-parameter combination that yields the best result. As a proof of principle we chose to maximise the product of the purity and completeness at a given SNR. Initially we chose SNR = $4$ but found that we could achieve perfect results there. As a result we maximised  $PC3$ (the product of purity and completeness at SNR = 3), as discussed in section (\ref{sec:pc}). This is a reasonable, but non-unique choice. Depending on the science one may wish instead to maximise purity at a fixed completeness or vice versa. Nothing would change in the formalism we present.  

\begin{figure*}
\begin{center}
\includegraphics[width=0.8\textwidth]{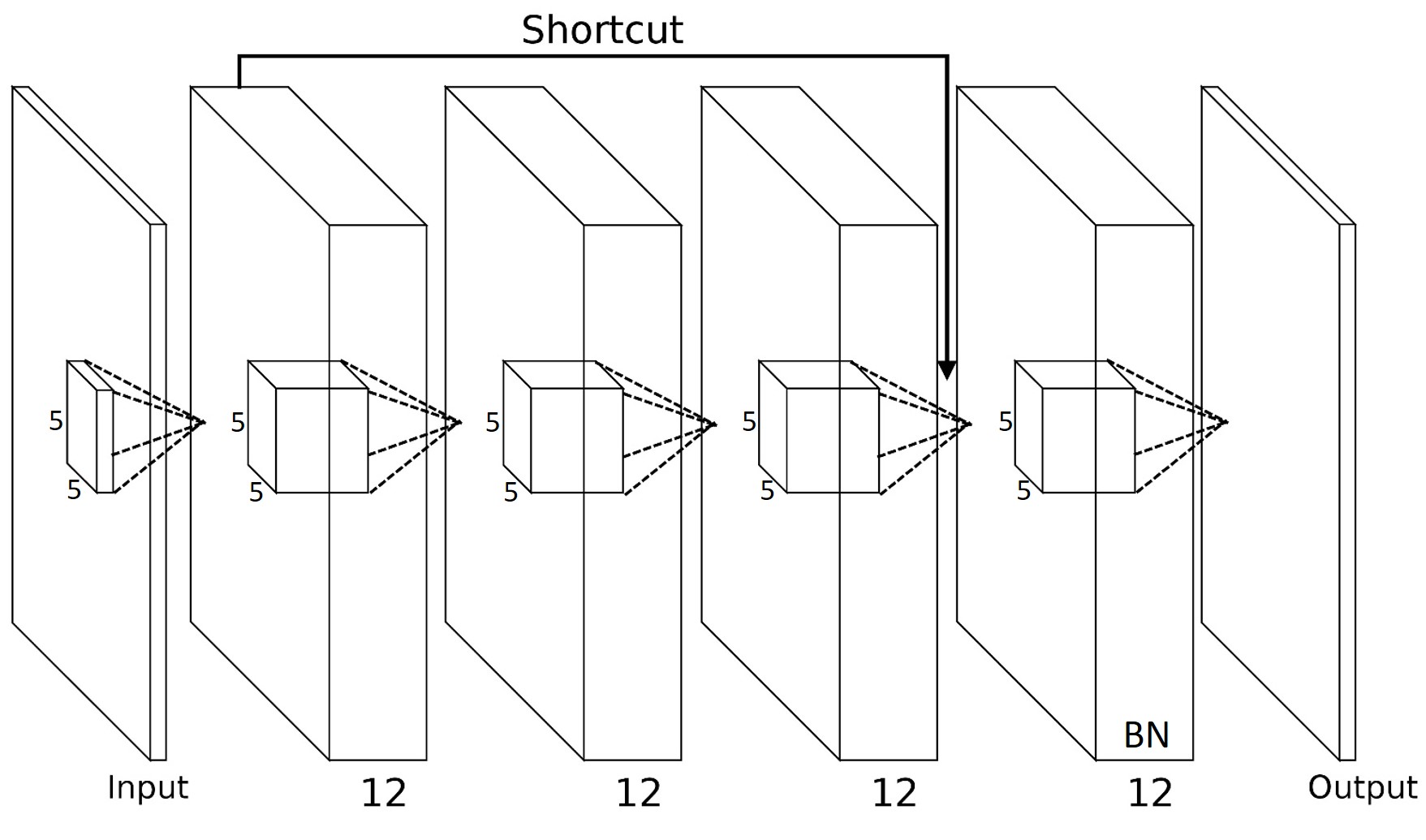}
\end{center}
\caption{The network architecture used in \name. It includes $5$ intermediate layers with RELU activations. Each layer has twelve $5\times 5$ filters while the last one reduces the $12$ filters to one. There is a shortcut and batch normalization after the third layer which helps deeper layers learn better.}
\label{network}
\end{figure*}

The hyper-parameters are chosen by splitting the data into training, validation and test sets to avoid over-fitting. Given the heavy computational cost of hyper-parameter optimization, it was not possible to survey all of the possible configurations of the parameters, though this would be sensible when training the algorithm for actual deployment. For example, the smoothing hyper-parameters were adjusted in the early stages of the analysis to find good values and then kept fixed. For the CNN architecture we played with different numbers of layers and filter sizes and again made a final choice early on. The best architecture is shown in Fig.~\ref{network} which includes $5\times 5$ filters and a batch normalization after a shortcut. 

The hyper-parameters we did vary systematically in this study were $(\alpha,b_n,\Delta_{max},A_{min})$, as well as the blob center extraction method. These were chosen by maximising $PC3$ on the validation set. We found that the best hyper-parameter configuration  is $\alpha=0.75$, $b_n=0.05$, $\Delta_{\rm max}=100$, $A_{\rm min}=10$ and Equation \ref{mean} for the point source extraction from blobs.

\section{Results}\label{Results}

In this section, we evaluate \name by comparing it against PyBDSF using the simulations described in Section \ref{Image Generation}.  The performance of both source-finding algorithms is assessed using purity (P) and completeness (C) as metrics, which are described in Section \ref{sec:evaluation}.

One potential advantage of a deep learning approach that is trained specifically for a given telescope is that it can learn to ignore the pattern of correlated noise in the images. We perform a preliminary investigation of this hypothesis by taking advantage of the fact that this noise pattern is dependent on sky position due to changes in the beam pattern. In Section \ref{sec:same_diff}, we test whether \name trained on the same field outperforms \name trained on different fields, when tested on simulated images from that given field.

Finally, the effects of using different weighting values in the $\it{uv}$ plane are investigated (see Section \ref{sec:weightings} for a detailed discussion on visibility weightings). We also investigate the effects of changing the crossmatching radius $R_{\rm cm}$, which can affect the results of any source detection algorithm.  For this study, $R_{\rm cm}$ is chosen to be $2/3$ of the synthesized beam size, unless specified otherwise, and this choice is motivated in Section \ref{sec:crossmatching}.

\subsection{Evaluating Algorithm Performance}
\label{sec:evaluation}
Purity is defined as the fraction of the total number of detected sources that are real, and is given by:
\begin{equation} \label{eq:purity}
Purity\left(f\right) = \frac{TP\left(f\right)}{TP\left(f\right) + FP\left(f\right)},
\end{equation}
where $TP\left(f\right)$ and $FP\left(f\right)$ denote the true positive and false positive counts, both of which are functions of flux (or alternatively SNR).  The TP counts refer to sources recovered by the source finder for which a simulated ground-truth source exists and is located within the crossmatching radius $R_{\rm cm}$.  False negative (FN) counts refer to detections from the source finder that cannot be matched with simulated ground-truth sources from the true catalogue within a radius $R_{\rm cm}$.

Completeness is defined as the fraction of actual ground-truth sources that are successfully recovered by the algorithm, and is given by:
\begin{equation} \label{eq:completeness}
\mbox{Completeness}\left(f\right) = \frac{TP\left(f\right)}{TP\left(f\right) + FN\left(f\right)},
\end{equation}
with $FN\left(f\right)$ denoting the false negative counts, i.e. the number of simulated ground-truth sources within a flux or SNR bin that the source finder failed to recover.

In order to establish whether or not detected sources can be classified as TP detections, positional crossmatching is performed between the simulated ground-truth catalogue and the source finder's catalogue.  If a supposed detection is located within a distance of $2/3$ of the synthesized beam size (referred to as the crossmatching radius $R_{\rm cm}$), where the beam size is given by Equation \ref{Beam_size}, the detection is labelled as a TP.  When multiple detections are located within a circle of radius $R_{\rm cm}$ centered on the ground-truth source, the detection closest to the considered ground-truth source is chosen as the correct TP. This avoids assigning multiple TP detections to a single ground-truth source.

\subsubsection{PC Score}
\label{sec:pc}
We also find it useful to multiply the purity and the completeness to create a ``PC'' score at different SNR values. For instance, while two algorithms may have the same PC score at an SNR of 4 (PC4), they may have very different PC3 values and the algorithm with the \emph{higher} PC3 would be more reliable at detecting fainter sources.

\subsubsection{S90 Score}
\label{sec:s90}
The true test of a source-finding algorithm is whether it can recover faint sources from an image. To evaluate this, we used a metric called S90, which is defined as the lowest SNR above which an algorithm (either PyBDSF or \name) produces both purity and completeness values of at least 0.9. In general it is expected that a point source detection algorithm with a \emph{lower} S90 score will be able to reliably produce a larger catalogue of faint sources, effectively increasing the sensitivity of the telescope.

\subsection{Comparison of \name and PyBDSF}
\begin{figure*}
\begin{center}
\includegraphics[width=1.0\linewidth]{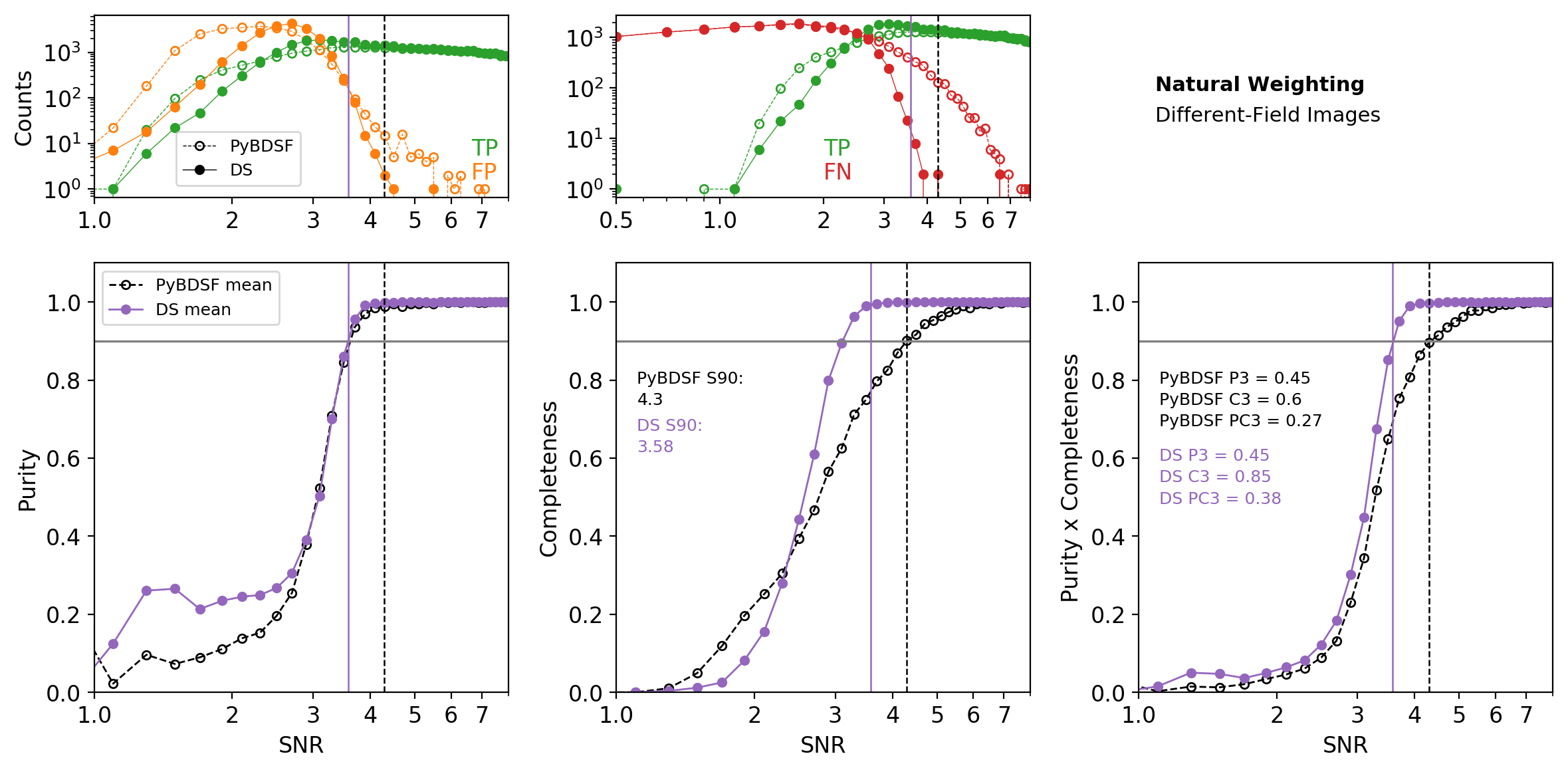}
\caption{{\bf Comparative performance on Natural Weighting images}: A comparison of the performance of \name (solid purple with filled circles) and PyBDSF (dashed black with open circles), using natural weighting in the $uv$-plane, applied to the different-field images. The bottom panels represent Purity (P, left), Completeness (C, center) and purity$ \times$ completeness (right) metrics, while the top panels represent the corresponding true positive (TP; green curves) false positive (FP; orange curves) and false negative (FN; red curves) counts used in calculating the purity and completeness curves. The $S90$ metric (the SNR at which both purity and completeness are at least 0.9, see Section \ref{sec:s90}) is marked by the vertical lines for PyBDSF (dashed black) and \name (solid purple), showing how \name allows to push to significantly fainter sources at the same PC threshold.}
\label{fig:ds9_vs_pybdsf_natural}
\end{center}
\end{figure*}

\begin{figure*}
\begin{center}
\includegraphics[width=1.0\linewidth]{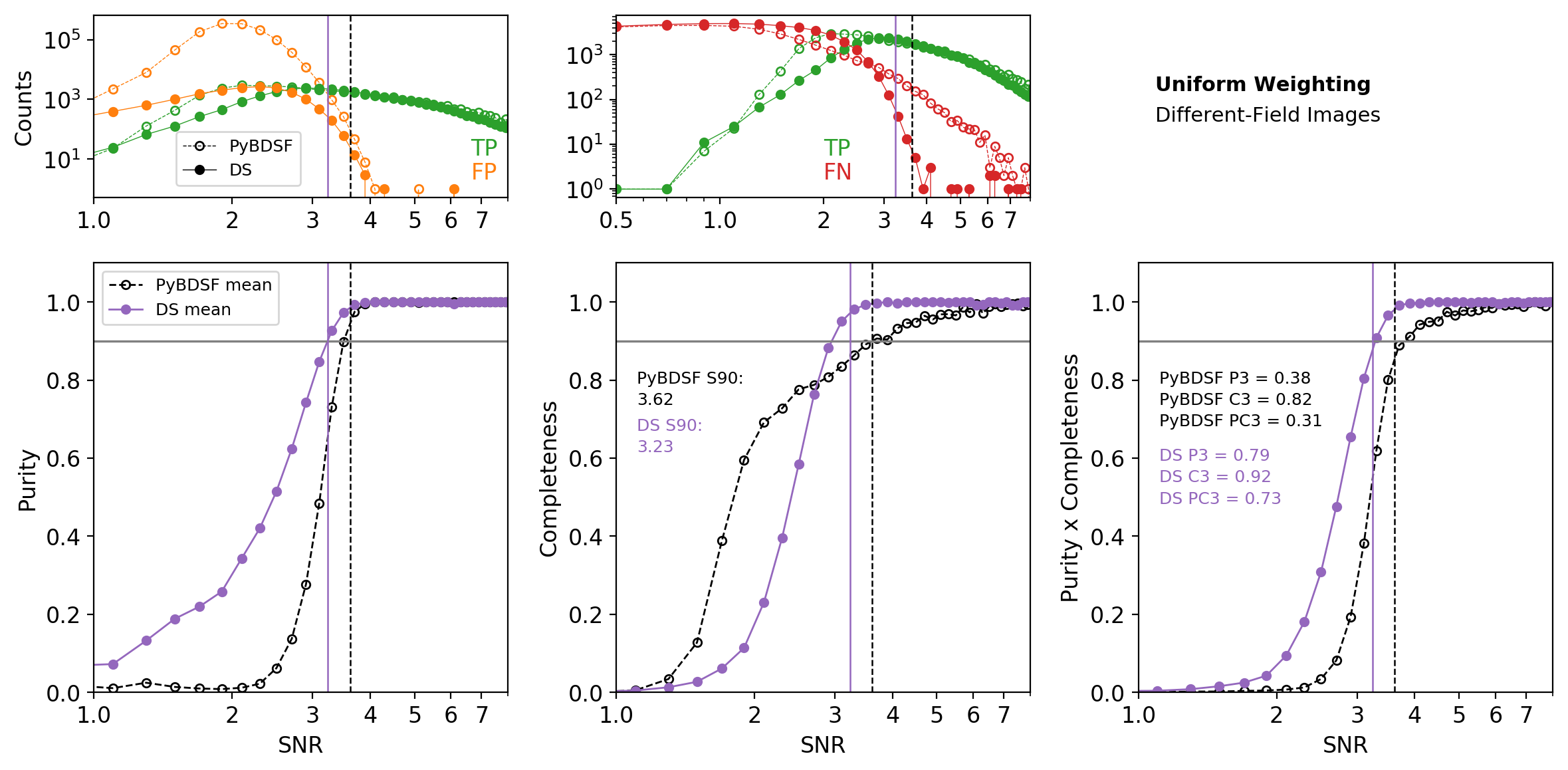}
\caption{{\bf Comparative performance on uniform-weighted images}: similar to Fig.~ \ref{fig:ds9_vs_pybdsf_natural}, of \name (solid purple with filled circles) and PyBDSF (dashed black with open circles), but this time using uniform weighting in the $uv$-plane.  The bottom panels represent metrics for purity (left), completeness (center) and purity $\times$ completeness (right), while the top panels represent the corresponding true positive (TP; green curves) false positive (FP; orange curves) and false negative (FN; red curves) counts used in calculating the purity and completeness curves.  The $S90$ metric (the SNR at which both purity and completeness are at least 0.9, see Section \ref{sec:s90}) is given by the vertical lines for PyBDSF (dashed black) and \name (solid purple).}
\label{fig:ds9_vs_pybdsf_uniform}
\end{center}
\end{figure*}

After performing crossmatching between the algorithm's output catalogue (either PyBDSF or \name) and the source catalogue (containing ground truth coordinates of the sources), we calculate the average purity and completeness curves for the test-set results, which consists of 290 images.

The bottom panels of Fig.~\ref{fig:ds9_vs_pybdsf_natural} compare the purity and completeness performance of PyBDSF to that of \name for images (created at a variety of sky positions) where natural weighting is applied to the $uv$-plane samples.  For this scenario, even though \name shows no improvement in purity above a SNR of 3, a significant improvement is observed in completeness.  At a given reference SNR of 3, \name achieves a $\sim40\%$ higher average completeness level ($0.85$) compared to PyBDSF ($0.6$), while both algorithms have similar purity values at this SNR ($0.45$).  As a result, when combining the two metrics \name gives a better overall purity $\times$ completeness performance than PyBDSF.

The S90 metrics (Section \ref{sec:s90}) for \name and PyBDSF are represented by the dashed black and solid purple vertical lines respectively in Fig.~ \ref{fig:ds9_vs_pybdsf_natural}, with the $90\%$ purity and completeness level highlighted by the horizontal grey lines.  For PyBDSF, both purity and completeness exceed the $90\%$ quality level at SNR values above S90 $= 4.3$.  \name on the other hand reaches this performance at a lower SNR of S90 $= 3.58$, which illustrates the ability of \name to accurately detect fainter point sources.

The top panels of Fig.~\ref{fig:ds9_vs_pybdsf_natural} give the true positive (TP; green lines), false positive (FP; orange lines) and false negative (FN; red lines) counts for each SNR bin (see Section \ref{sec:evaluation} for definitions), integrated over the 290 test images.  

Interestingly, both \name (solid lines) and PyBDSF (dashed lines) have similar TP counts above SNR of $\sim4$ for the range shown here, but \name has much fewer FP detections compared to PyBDSF.  In addition, it is evident from the red curves that \name has fewer FN detections (i.e. missed sources) than PyBDSF above SNR of $\sim3$.  Even though PyBDSF has more TP detections than \name below SNR of $\sim2.3$, \name has many fewer FP counts compared to PyBDSF, making it a more reliable algorithm for faint sources.

\subsection{Natural vs. Uniform Weighting}
From Fig.~\ref{fig:ds9_vs_pybdsf_uniform}, similar qualitative results are obtained for the case when uniform weighting is applied to visibility samples in the $uv$-plane.  Here \name achieves optimal purity and completeness performance (i.e. above the $90\%$ quality level) at SNR$=3.23$, while PyBDSF achieves this performance at a higher SNR of 3.62.  In general, \name shows higher purity performance than PyBDSF at all SNR, whereas for completeness, \name only exceeds PyBDSF below SNR of $\sim2.8$.  Combining the purity and completeness metrics, 
it can be seen that \name has overall better PC performance than PyBDSF at all SNR, which shows that \name is also able to reliably recover fainter sources than PyBDSF for the uniform weighting scenario.  

For easy reference, Table \ref{table:pcvalues} gives a concise summary of the purity (P), completeness (C) and purity $\times$ completeness (PC) performance for \name and PyBDSF, as given in Figs~\ref{fig:ds9_vs_pybdsf_natural} and 
\ref{fig:ds9_vs_pybdsf_uniform}.  Numerical values for these metrics are quoted at SNR values of 3, 4 and 5, where bold fonts indicate the best performing algorithm for a given SNR.

As previously mentioned, a lower S90 score for an algorithm translates into it being able to reliably recover fainter sources.  As a result, it is generally expected that such an algorithm will produce catalogues with a higher number of source counts.  This was verified for \name and PyBDSF by integrating the TP and FP counts over all the SNR bins above S90.  Table \ref{table:counts} gives the S90 values of each algorithm for both natural and uniform weighting, along with the corresponding catalogue source counts where purity and completeness are above $0.9$.  The bold numbers show that \name consistently yields $\sim$ 10\% more sources for both the natural and uniform scenarios. However, the exact number of sources detected depends entirely on the underlying flux distribution of sources, which we have chosen somewhat arbitrarily, so it is possible that the expected increase in detected sources could be higher for a more realistic distribution.

\begin{table}
\begin{center}
	\begin{tabular}{c|ccc|ccc}
    	\hline\hline
        {} & \multicolumn{6}{c}{Natural Weighting} \\
        \hline
        {} & \multicolumn{2}{c}{P} & \multicolumn{2}{c}{C} & \multicolumn{2}{c}{PC}\\
        \hline
         SNR & DS & PyB & DS & PyB & DS & PyB\\
        \hline
        3 & {\bf 0.45} & {\bf 0.45} & {\bf 0.85} & 0.60 & {\bf 0.38} & 0.27 \\
        4 & {\bf 0.99} & 0.98 & {\bf 1.00} & 0.85 & {\bf 0.99} & 0.83 \\
        5 & {\bf 1.00} & 0.99 & {\bf 1.00} & 0.96 & {\bf 1.00} & 0.95\\
        \hline\hline
        {} & \multicolumn{6}{c}{Uniform Weighting} \\
        \hline
        {} & \multicolumn{2}{c}{P} & \multicolumn{2}{c}{C} & \multicolumn{2}{c}{PC}\\
        \hline
         SNR & DS & PyB & DS & PyB & DS & PyB\\
        \hline
        3 & {\bf 0.79} & 0.38 & {\bf 0.92} & 0.82 & {\bf 0.73} & 0.31 \\
        4 & {\bf  1.00} & 1.00 & {\bf 1.00} & 0.92 & {\bf 1.00} & 0.92 \\
        5 & {\bf1.00} & 1.00 &{\bf 1.00} & 0.96 &{\bf 1.00} & 0.96 \\
        \hline\hline
	\end{tabular}   
\end{center}
\caption{Purity (P), Completeness (C) and Purity $\times$ Completeness (PC) values (Section \ref{sec:evaluation}) for PyBDSF (PyB) and \name (DS) at SNR values of 3, 4 and 5, showing an improvement of \name over PyBDSF of $40\%$ for natural weighting and $120\%$ for uniform weighting, in PC at SNR = 3. P, C and PC values for the natural-weighted images are shown in the upper panel of the table (corresponding to results from Fig.~\ref{fig:ds9_vs_pybdsf_natural}), while metrics in the lower panel correspond to uniform weighting in the $uv$-plane (according to Fig.~\ref{fig:ds9_vs_pybdsf_uniform}).\label{table:pcvalues}}

\end{table}

\begin{table}
\begin{center}
	\begin{tabular}{ccc|cc}
    	\hline\hline
        {} & \multicolumn{2}{c}{PyBDSF} & \multicolumn{2}{c}{DS9$_D$}\\
        \hline
        {} & counts & S90 & counts & S90\\
        \hline\hline
        Natural weighting & 50\,545 & 4.30 & {\bf54\,895} & {\bf3.58} \\
        Uniform weighting & 17\,294 & 3.62 & {\bf19\,857} & {\bf3.23} \\
        \hline\hline
	\end{tabular}   
\end{center}
\caption{Catalogue counts (obtained by summing the TP and FP counts) for PyBDSF and \name, integrated over all SNR bins above the S90 quality metric for natural and uniform $uv$-plane weighting scenarios.  These counts correspond to the TP and FP detections of the respective algorithms for all of the 290 test-phase images, each of which contains 300 sources at different SNR values.  Both PyBDSF and \name were applied to (and trained on) images simulated at different locations in the sky.  On average, it is expected that an algorithm with a lower S90 metric will produce a larger catalogue of detections while still adhering to the constraint that both purity and completeness levels remain above the $90\%$ quality level.  Bold-face fonts highlight the best performing algorithm in each case in terms of both the S90 metric and the number of counts. \label{table:counts}}
\end{table}
\begin{figure*}
\begin{center}
\includegraphics[width=\linewidth]{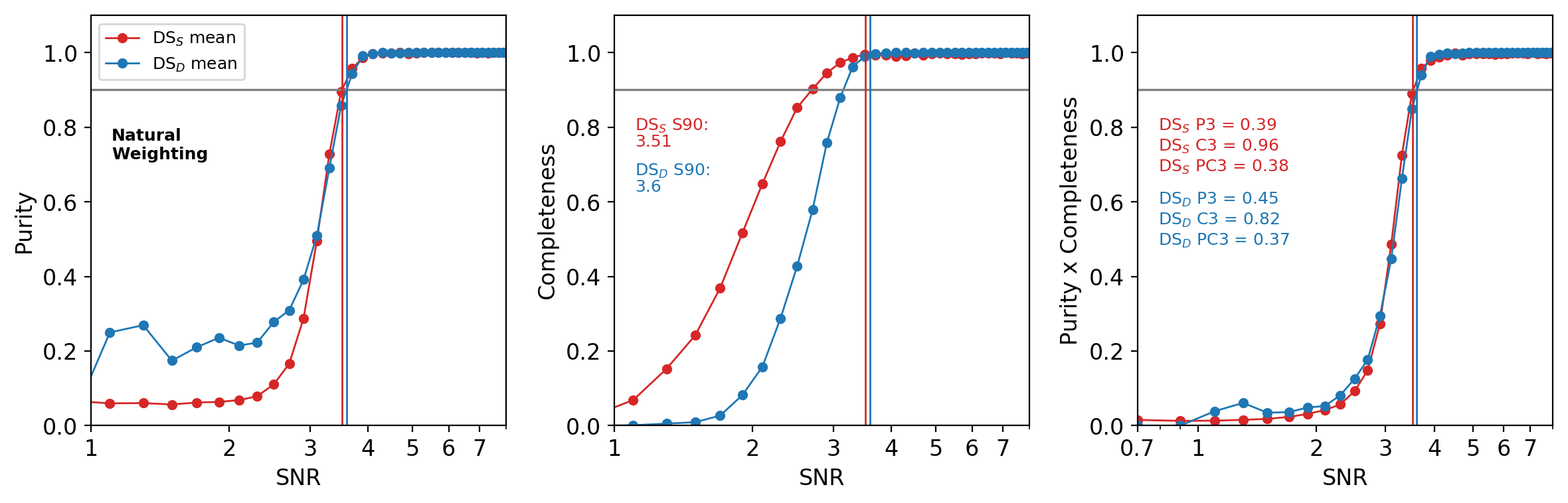}\\
\includegraphics[width=\linewidth]{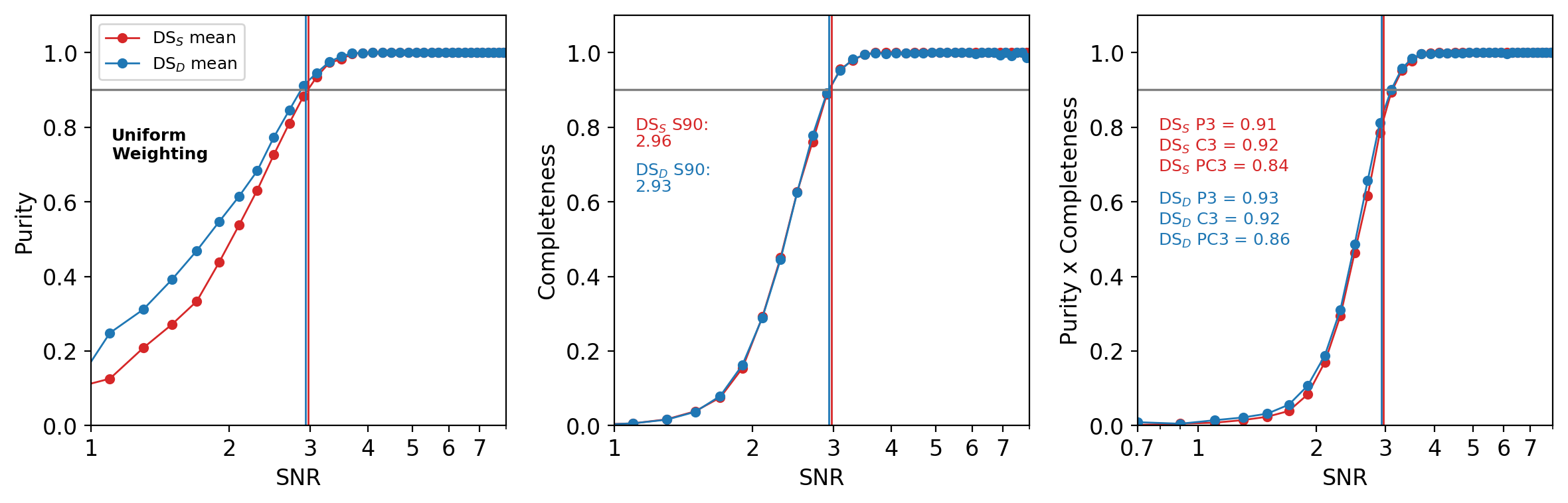}
\caption{{\bf Comparison of Different vs Same field performance}: The purity (P), completeness (C) and purity $\times$ completeness (PC) curves for \name, where performance results are compared for when training was done on same-field images (\namesame; red curves) and different-field images (\namediff; blue curves). At least in the cases simulated there is no advantage to training on simulations specific to the test data.  The top and bottom panels represent results from images produced using natural and uniform weighting in the $uv$-plane respectively.  The horizontal gray lines highlight the $90\%$ quality level, while the vertical red and blue lines correspond to the S90 metrics for \namesame ~and \namediff ~respectively (Section \ref{sec:s90}).  For all the cases considered here, \name was applied to same-field images at test time.  \label{fig:same_vs_diff_training}}
\end{center}
\end{figure*}

\subsection{Training on Same-Field and Different-Field Images}
\label{sec:same_diff}

As discussed in Section \ref{sec:DeepSource}, \name is a deep CNN that is trained on ground truth data with the goal of being able to generalize to new unseen data.  As part of assessing the inference performance of \name, training was performed on both same-field (\namesame) and different-field (\namediff) images separately, where each trained model was evaluated using same-field images. Same-field images refer to images that are produced from simulations of the same region of sky, i.e. the field-centers of images are kept constant.  Conversely, different-field images refer to images produced by simulating different regions of the sky (see Sections \ref{Model Simulation} and \ref{Image Generation} for in-depth discussions on how simulations were set up for this study).  

In the following section, \namediff ~refers to when \name has been trained on different-field images, while \namesame ~refers to when \name has been trained on same-field images.  The aim with this approach is to test whether the algorithm can learn local noise structure of a particular sky position.

Fig.~\ref{fig:same_vs_diff_training} shows the purity and completeness curves for \name trained on same-field and different-field images, where performance was evaluated on same-field images at test time.  Training on same-field images seem to improve completeness performance at the expense of lower purity below SNR of ~3 for natural-weighted images (top panel).  As a result, very little difference is observed in the PC performance.  
For the uniform-weighting scenario, both \namesame ~and \namediff ~have similar completeness performance, where \namesame ~has marginally lower purity performance.

Even though \namesame ~seems to improve certain metrics when performing detection at test-phase, results from training on same-field images remain inconclusive.  Training on same-field images was therefore not adopted as the optimal training strategy. This result is surprising, however more testing of multiple fields would be required to confirm that it is generally true that training on a given field does not improve performance for that field. This will be done in future work.

\subsection{Effects of Different Crossmatching Radii}
\label{sec:crossmatching}
The crossmatching radius, $R_{\rm cm}$, defines the maximum distance that a detected source can be from a true source for them to be associated and assigned a true positive detection label.  In order to motivate our choice of using a crossmatching radius equal to two-thirds of the synthesized beam for this study, we compare how the PC curves for PyBDSF and \name change with different values of $R_{\rm cm}$ when applied to different-field images.

\begin{figure}
\begin{center}
\includegraphics[width=0.9\linewidth]{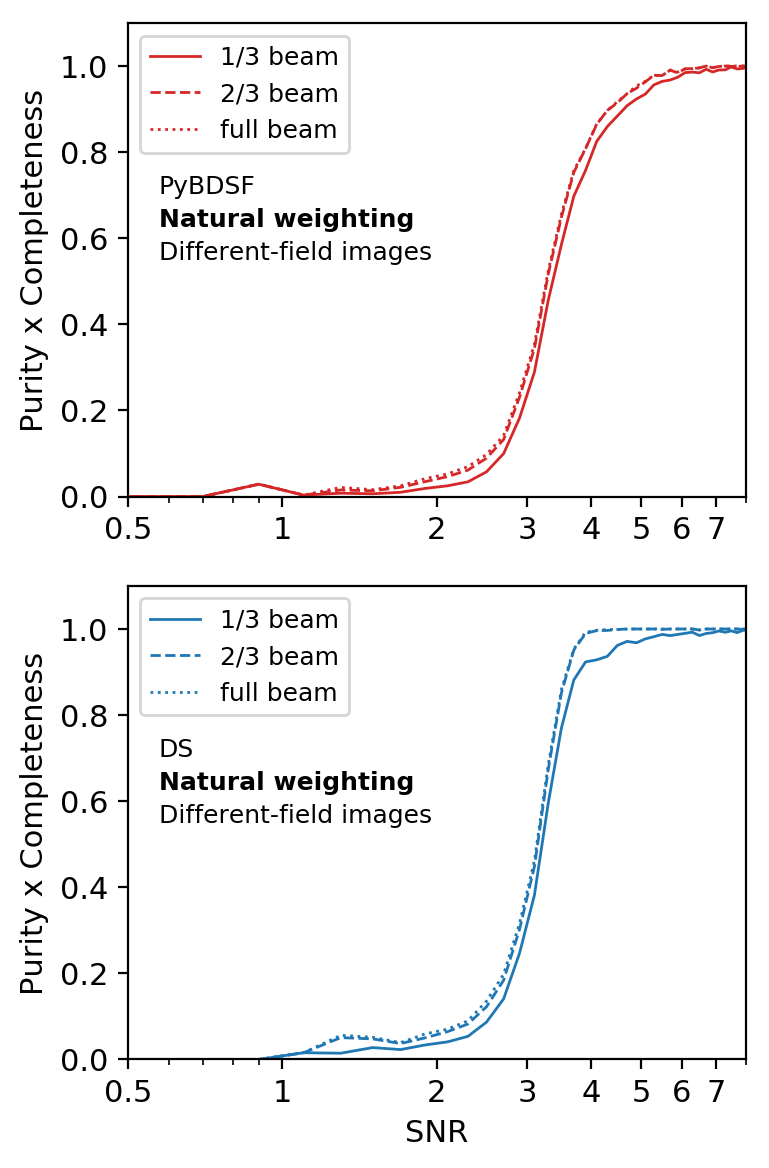}
\caption{A comparison of the purity $\times$ completeness (PC) curves for different values of the crossmatching radius ($R_{\rm cm}$), namely $1/3$ (solid curves), $2/3$ (dashed curves) and full beam sizes (dotted curves).  The top and bottom panels correspond to PyBDSF and \name results respectively, where both algorithms were applied to different-field images at test time. In this paper we chose $2/3$ for $R_{\rm cm}$ but it is clear the results show only weak dependence on $R_{\rm cm}$. \label{xm_radius_comparison}}
\end{center}
\end{figure}

Fig.~\ref{xm_radius_comparison} shows the PC curves for PyBDSF (red) and \name (blue).  Here $R_{\rm cm}$ takes on values of one-third (solid lines), two-thirds (dashed lines) and full (dotted line) beam size, where we now only consider images from the naturally-weighted $uv$-plane samples.  

Both \name and PyBDSF are affected similarly when changing the cross-matching radius.  The largest increase in performance occurs when changing $R_{\rm cm}$ from $1/3$ to $2/3$ of the beam size for both \name and PyBDSF.  Almost no improvements are observed when increasing $R_{\rm cm}$ from $2/3$ to a full beam size.  In addition, \name improves noticeably around a SNR of 4.

The small variability when changing $R_{\rm cm}$ is a good indication that our choice for $R_{\rm cm}$, along with the subsequent crossmatching, does not affect the final results but produces unbiased purity and completeness metrics. In a more general analysis $R_{\rm cm}$ can be made a hyper-parameter to be optimised. 

\subsection{Comparison of DS9 Training Times}
\begin{figure*}
\begin{center}
\includegraphics[width=1.0\linewidth]{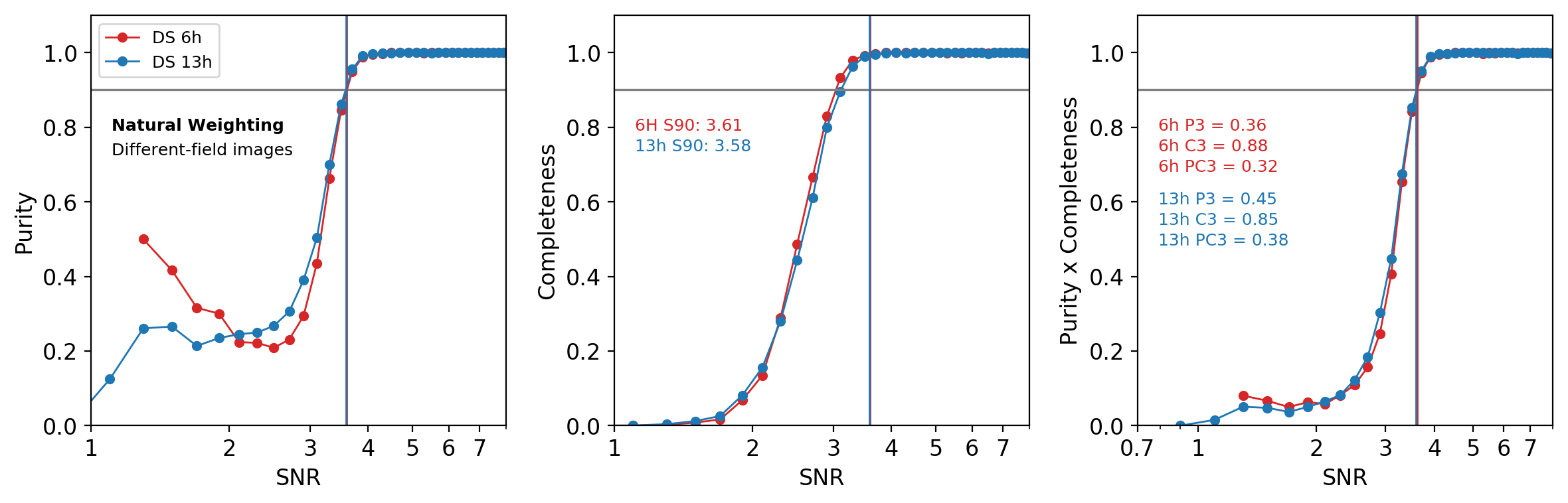}
\includegraphics[width=1.0\linewidth]{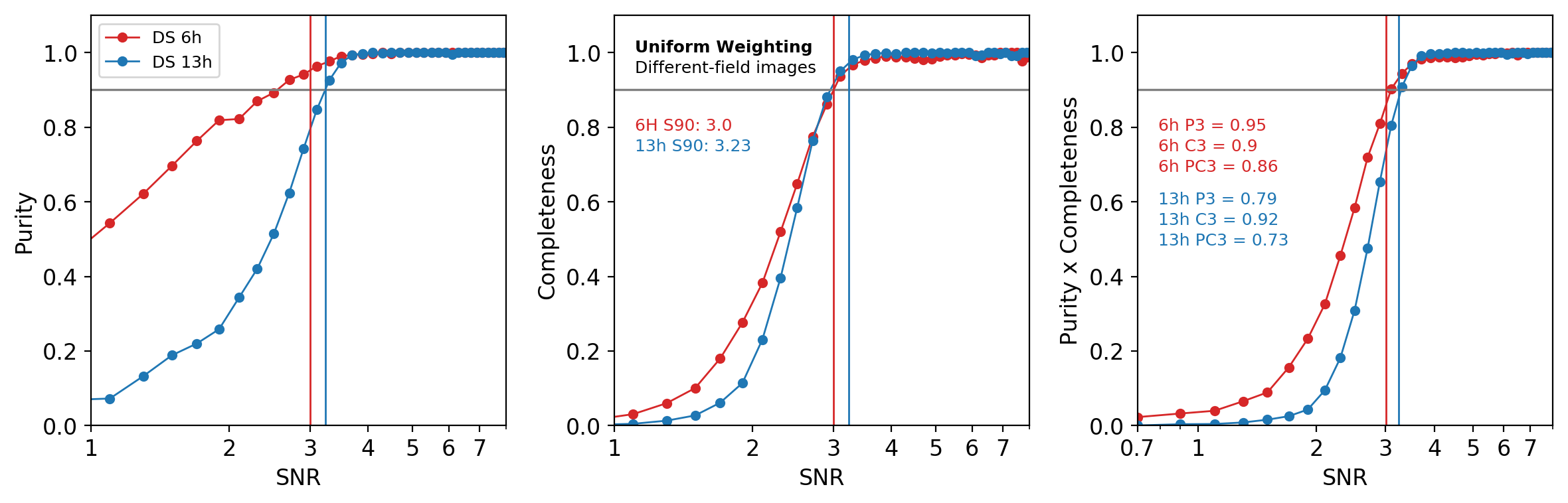}
\caption{{\bf Effect of training time:} Shown here are the purity (P), completeness (C) and purity $\times$ completeness (PC) metrics for \name trained on different-field images for  obtained when using natural weighting in the $uv$-plane.  The red curves represent the performance of \name after training for a time of 6 hours, while the blue curves represent the performance after 13 hours of training.  The horizontal gray lines indicate the $90\%$ quality level in the metrics, while the vertical red and blue lines correspond to the S90 metrics for the 6 hour and 13 hour training times respectively. The results show that longer training times do not lead to improved performance of \name for the cases considered here, and all results shown in this paper are for 6 hours of training.}
\label{training_time_comparison}
\end{center}
\end{figure*}

One question one might have is whether training for longer yields better results. In Fig.~\ref{training_time_comparison} we show the results for training \name for 6 hours (red curves) compared to 13 hours (blue curves).  For the natural weighting case (top panels), there is essentially no improvement when training for longer.  Consequently, the S90 metric for both the 6 hour and 13 hour training times are very similar: 3.61 and 3.58 respectively.

When considering uniform weighting (bottom panels), there is a notable difference between results produced after training for 6 hours and for 13 hours.  In most cases it is expected that training for longer will improve inference performance, however, for the uniform weighting scenario the longer training time doesn't improve purity and completeness performance.  This suggests that 6 hours could be sufficient to maximize performance for the case considered, but the performance may decrease with longer training time due to overfitting noisy images. This is a topic for further investigation in the future.

\section{Conclusions}

Extraction of point sources is an important task in many fields of vision and key to astronomy. Here we apply deep Convolutional Neural Networks (CNNs) to the detection of point sources in interferometric radio astronomy images and compare the resulting performance against a standard code in radio astronomy: PyBDSF.  Given the excellent performance of CNN in general vision problems one expects a CNN approach to be able to learn the correlated noise structure of such images allowing improved sensitivity and a smaller false positive and false negative rates. We indeed find that our CNN implementation - which we dub \name - outperforms a state-of-the-art source detection algorithm, PyBDSF, in all metrics. 

\name is trained on simulated maps with known point source positions and learns to amplify the Signal-to-Noise (SNR) of the input maps. We then convert this map to a catalogue of sources using a dynamic blob detection algorithm. In each case (Different vs Same and natural weighting vs uniform) we trained \name on 200 simulated MeerKAT images, optimised hyper-parameters on a validation set of 10 images, and tested on the remaining 290 images.  Each of these images contains 300 sources of which $\sim 45\%$ are faint, with SNR < 1.

We compare the performance of \name on the test images to the widely-used code PyBDSF, optimised on the validation set and applied to the same test images. An advantage of PyBDSF is that it requires no training, but this also means there is   little scope to fine-tune it for specific surveys (limited to just two thresholds). To perform the comparison we computed the purity (P; the fraction of claimed point sources that really are point sources) and the completeness (C; the fraction of point sources that are correctly detected) as a function of SNR. We typically find that while the purity is similar for both algorithms, the completeness is significantly higher for \name. For natural weighting \name achieves at least 90\% in both purity and completeness (a quantity we call S90) down to a SNR of 3.6 compared with only 4.3 for PyBDSF, while for uniform weighting the corresponding figures for \name (PyBDSF) are 3.2 (3.6). The result of this is that for the same observation, our trained CNN can detect more faint sources than a standard source-finding algorithm like PyBDSF at the same purity. 

It might be expected that if the CNN is trained on simulations of the same field and then tested on that same field (defined by a specific RA and DEC), it should outperform a CNN trained on a variety of different fields. This is because the beam changes with sky position and we'd expect the CNN to be able to learn the exact noise pattern of that field. However when we tested this, we found the performance is surprisingly similar between the same/different fields. This may be due to our particular choice of field (which was close to equatorial where the beam is quite circular) or perhaps the CNN is able to learn the noise pattern without requiring multiple simulations of a given field. We leave it to future work to test this and other extensions further.

Finally we note that it is likely that there are even better deep learning implementations for point source detection than the one presented here. It is also important to bear in mind that any supervised machine learning algorithm is only as good as the training data used. Since the underlying simulations used to train the algorithms will never perfectly match reality, there will likely always be a useful role for robust algorithms like PyBDSF that require little or no training. Beyond point source detection we also want an algorithm that detects extended sources and diffuse emission. We leave this important extension to future work.  The \name code is available at [{\em link available on acceptance of paper}]. 

Note added: In the final stages of finishing this work a paper was released by \cite{wu2018} on similar topics.

\section*{Acknowledgements}
We thank Yabebal Fantaye, Mat Jarvis, Marzieh Farhang, Russ Taylor and Mattia Vaccari for discussions. We acknowledge support from the STFC through the Newton Fund, SKA Africa and the National Research Foundation (NRF) towards this research. Opinions expressed and conclusions arrived at, are those of the authors and are not necessarily to be attributed to the NRF.  The numerical simulations were carried out on Baobab at the computing cluster of University of Geneva and on compute facilities at SARAO.



\bibliographystyle{mnras}
\bibliography{reference}



\bsp	
\label{lastpage}
\end{document}